\documentclass{article}

\usepackage{arxiv}

\usepackage[utf8]{inputenc} 
\usepackage[T1]{fontenc}    
\usepackage{hyperref}       
\usepackage{url}            
\usepackage{booktabs}       
\usepackage{amsfonts}       
\usepackage{nicefrac}       
\usepackage{microtype}      
\usepackage{cleveref}       
\usepackage{lipsum}         
\usepackage{graphicx}
\usepackage[square,sort,comma,numbers]{natbib}
\usepackage{doi}
\DeclareUnicodeCharacter{2212}{-}

\title{Active Control of Flow over Rotating Cylinder by Multiple Jets using Deep Reinforcement Learning}

\author{{Kamyar Dobakhti} \\
	Mechanical Engineering Department\\
	University of Zanjan\\
	Zanjan, Iran\\
	\And
	{Jafar Ghazanfarian} \\
	Mechanical Engineering Department\\
	University of Zanjan\\
	Zanjan, Iran \\
	\texttt{j.ghazanfarian@znu.ac.ir} \\
}

\renewcommand{\shorttitle}{\textit{arXiv} Template}

\hypersetup{
pdftitle={Active Control of Flow over Rotating Cylinder by Multiple Jets using Deep Reinforcement Learning},
pdfauthor={Kamyar Dobakhti, Jafar Ghazanfarian},
}

\begin{document}
\maketitle

\begin{abstract}
	The real power of artificial intelligence appears in reinforcement learning, which is more sophisticated due to its dynamic nature. Rotation and injection are some of the proven ways in active flow control for drag reduction on blunt bodies. In this paper, rotation will be added to the cylinder alongside the deep reinforcement learning (DRL) algorithm, which uses multiple controlled jets to reach the maximum possible drag suppression. Characteristics of the DRL code, including control parameters, their limitations, and optimization of the DRL network for use with rotation will be presented. This work will focus on optimizing the number and positions of the jets, the sensors location, and the maximum allowed flow rate to jets in the form of the maximum allowed flow rate of each actuation and the total number of them per episode. It is found that combining the rotation and DRL is promising since it suppresses the vortex shedding, stabilizes the Karman vortex street, and reduces the drag coefficient by up to $49.75\%$. Also, it will be shown that having more sensors at more locations is not always a good choice and the sensor number and location should be determined based on the need of the user and corresponding configuration. Also, allowing the agent to have access to higher flow rates, mostly reduces the performance, except when the cylinder rotates. In all cases, the agent can keep the lift coefficient at a value near zero, or stabilize it at a smaller number.
\end{abstract}

\maketitle

\section{Introduction}
Active flow control is a long-standing topic in fluid mechanics. By using different kinds of actuators, it alters the flow behavior to improve the aerodynamics/hydrodynamic performance of a flow system. One such usage is for drag reduction, which is of importance because annually a large amount of energy is being consumed to overcome the drag forces encountered in many engineering applications such as ground, air, and sea vehicles \cite{M.Gad-el-Hak}, For example, in ground vehicles, the aerodynamics drag has a part of approximately $27\%$ in the total fuel consumption \cite{Kornilov}. Thus, achieving effective and stable flow control for drag force reduction is important. Unfortunately, due to the combination of non-linearity, time-dependence, and high-dimensionality intrinsic to the Navier-Stokes equations and the computational power needed for calculations, finding efficient strategies for performing active flow control is highly sophisticated \cite{Brunton,Duriez}.

Active flow control has been discussed in simulations using reduced-order models and harmonic forcing \cite{Bergmann}, direct simulations coupled with the adjoint method \cite{Flinois}, or linearized models \cite{Lee}, and mode tracking methods \cite{Queguineur}. Realistic actuation mechanisms such as acoustic excitation \cite{Kim}, synthetic jets \cite{Kim}, plasma actuators \cite{Zhang,Sato}, suction mechanisms \cite{Wang}, transverse motion \cite{Li}, periodic oscillations \cite{Lu}, oscillating foils \cite{Bao}, air jets \cite{Zhu}, and the Lorentz forces in conductive media \cite{Zhu}, are also discussed in detail, as well as limitations imposed by real-world systems\cite{Belson}.

Similar experimental work has been performed, to control either the cavitation instability \cite{Che}, the vortex flow behind a conical fore-body \cite{Meng}, or the flow separation over a circular cylinder \cite{Jukes}. While most of the literature has focused on complex, closed-loop control methods, some open-loop methods also have been discussed, both in simulations \cite{Meliga} and in experiments \cite{Shahrabi}. Despite the remarkable results from active flow control, achieving effective and stable flow control strategies is a highly sophisticated process, while there are candidates for both sensors and actuators to be used in such systems as the discussed papers have shown, finding algorithms for using those measurements and possible actuations to perform active flow control efficiently and effectively, is challenging and resources heavy. In addition, challenges such as disturbances inherent to real-world applications, imperfections in the building of the sensors or actuators, and adaptivity to the ever-changing external conditions are also parts of the problem.

The machine learning algorithms offer a completely different paradigm from the old active flow control methods. There are three different machine learning sub-domains, namely, supervised learning which uses historically labeled data, unsupervised learning which uses historically unlabeled data to find the pattern of the said data, and reinforcement learning which typically has a very different approach compared to the supervised and unsupervised learning as it does not rely on a historical data. Instead, it relies on a system of an agent, environment, observations, and rewards to teach itself the best way to accomplish a task through self-exercise.

In a typical reinforcement learning algorithm, there is an agent that works based on a specific policy, interacts with an environment in a closed-loop fashion, and receives a reward corresponding to an action that the agent has taken in/on that environment. Then the agent learns from experiences/rewards to take actions that maximizes the expected result/cumulative reward. This means that it can reach quicker solutions without prior information about the physics of the problem \cite{Brunton}. Since the data-driven and the learning-based methods are suitable to be used in non-linear, high-dimensional, and complex problems, they are therefore, also suitable for performing active Flow Control \cite{Duriez}. More specifically, such promising methods include deep reinforcement learning.

In 2018 three papers\cite{Verma, Rebault19, Dressler} showed the success of the DRL method in performing AFC through using recurrent neural networks (RNN), deep neural networks (DRL) which could reduce drag coefficient around a cylinder using two jets by $8\%$, and deep Q-networks which used DRL to approximate Q-function and was able to control the flow in micro-channels as good as a human operator. In 2019 one paper\cite{Jean} used a multi-environment approach to speed up DRL learning and enable it to use more than one core per learning. Another paper\cite{Novati} used DRL to achieve gliding with either minimum energy expenditure or the fastest time of arrival, at a predetermined location. They found that model-free reinforcement learning leads to more robust gliding than model-based optimal control strategies with a modest additional computational cost. They also demonstrate that the gliders with DRL can generalize their strategies to reach the target location from previously unseen starting positions. In 2020, Hongwei et al.\cite{Hongwei}, used the same method for controlling four jets to reduce the drag force around a cylinder. Reynolds numbers of 100, 200, 300, and 400 were used, and they showed that the DRL-controlled jets can reduce the drag coefficient by $5.7\%$, $21.6\%$, $32.7\%$, and $38.7\%$, respectively. Another work\cite{Xu} used DRL to control two rotating cylinders behind a main cylinder as actuators, in a 2D flow field with a Reynolds number of 240, for drag reduction. They reported that the counter-rotating small cylinder pair were able to stabilize the periodic shedding of the main cylinder wake. Also, there have been cases that used RL for active flow control in turbulent flows, such work\cite{Dixia} was done for Re number of 10,160, and for flow control of a bluff cylindrical body in cross-flow, using two small rotating control cylinders in both experimental and simulation environments. The agent was able to reduce drag force by up to $30\%$ or reach another specified optimum point. Although it should be noted that in simulations, the current bottleneck is the computation limitations, as it would take a 64 cores Intel E5-2670 more than three weeks to finish 500 episodes. Another example is FengRen et al.\cite{FengRen} which used DRL for controlling jets around a cylinder to reduce the drag force in a flow with a Reynolds number of 1000 in simulation. Also, they were able to achieve a drag reduction of around $30\%$. In 2021 another paper\cite{Paris} focused on introducing the S-PPO-CMA algorithm which it's usage was to discard unnecessary or redundant sensor information.

Zheng et al.\cite{Changdong} compared the active learning framework to the reinforcement learning framework for suppressing vortex-induced vibrations. They showed that active learning can reduce the vibration amplitude of the cylinder by $28.33\%$, but RL can reduce it by $82.7\%$. Another work used DRL to achieve hydrodynamic stealth around bluff bodies \cite{Feng}. They used a group of windward-suction-leeward-blowing (WSLB) actuators and were able to reduce the deficit in streamwise velocity by $99.5\%$. There have been instances where DRL was used for open-loop control, such example is Ghraieb et al.\cite{Ghraieb}, They introduced a single-step proximal policy optimization (PPO)~\cite{Schulman}, a “degenerate” version of the PPO algorithm, intended for situations where the optimal policy to be learned by a neural network does not depend on state, as is notably the case in open-loop control problems. The approach proved relevant to map the optimum positions for placement of a small control cylinder in an attempt to reduce drag in laminar and turbulent cylinder flows. It was able to reduce the drag force of a square cylinder at Reynolds numbers in the range of a few thousand by $30\%$.

In 2022 a paper\cite{Wang} used DRL for AFC to control disturbed flow over a NACA 0012 airfoil under weak turbulent conditions of Re = 3000. When using constant inlet velocity, the agent was able to reduce drag by $27\%$ and enhance lift by $27.7\%$. Later, the pulsation at two different frequencies and their combination were applied for inlet velocity conditions, where the airfoil wake became more difficult to suppress dynamically and precisely; the reward function additionally contained the goal of saving the energy consumed by the jets, in this case, the DRL agent still was able to find a proper control strategy, where significant drag reduction and lift stabilization were achieved, and the agent was able to save the energy consumption of the synergetic jets by $83\%$. Yu-Fei et al.\cite{Mei} used DRL for performing AFC on ellipse, square, hexagon, and diamond models in laminar flow with Re number of 80 and were able to reduce drag by $11.50\%$, $10.56\%$, $8.35\%$, and $2.78\%$, (with 0 as the angle of attack) respectively. They also tested different angles of attack for ellipse, in which case the DRL agent was able to reduce the drag force for 5°, 10°, 15° and 20° AOA by $5.44\%$, $0.59\%$, $11.67\%$, and $0.28\%$, respectively. Tests were done for Reynolds numbers 160 and 320 with AOA of 0, in this case, the drag force reduction reached $10.84\%$ and $23.63\%$, respectively.

Deep reinforcement learning is also used for shape optimization. One notable paper\cite{Ren} used DRL for shape optimization as their main focus and introduced the SL-DDPG algorithm which was able to beat other trail-and-error methods and achieve a lift-to-drag ratio of 3.53. There are more papers on shape optimization which were published in 2021\cite{Jonathan} and 2022\cite{Hadi}. In 2022 Xie et al.\cite{Xie} used DRL for actively controlling heaving plate breakwaters and Hardman et al.\cite{Hardman} used DRL for the Manipulation of free-floating objects using Faraday flows in an experimental environment.

All papers showed a good drag reduction and all of this is possible because a breakthrough in RL research has been achieved after the integration between reinforcement learning and deep neural networks (DNNs), for what is called deep reinforcement learning (DRL). Artificial Neural networks are inspired by a simplification of biological neurons in an attempt to reproduce in machines some of the features that are believed to be at the origin of the intelligent thinking of the brain, i.e. biological neuron through mathematics, perceptron models \cite{LeCun}, and it is called the deep neural network when two or more of such networks are connected.

The main concept involves conducting computations through a group of basic processing units known as artificial neurons. Each neuron's output value is calculated by a non-linear function of the sum of its inputs. The connection between each neuron is called the edge. Neurons and edges typically have a weight, and as learning progresses, this weight is adjusted through an algorithm. This weight changes the strength of the signal at each edge. As an example, in supervised learning, these weights are tuned using algorithms such as stochastic gradient descent in order to minimize the cost function \cite{Goodfellow}. Considering the effectiveness of this method, It is theoretically possible for artificial neural networks to solve any problem as they are universal approximators. A feed-forward neural network with a non-linear activation function can fit any function with high accuracy \cite{Hornik}. This means that ANNs have the potential to be applied to virtually any problem or phenomenon that can be represented mathematically. However, the challenge in designing the ANNs, as well as developing the algorithms that train and utilize them, remains. The mentioned problems are currently an active area of research.

In the present work, we apply the deep reinforcement learning algorithm to an active flow control problem. The proximal policy optimization (PPO) method was used together with two 512 fully Connected artificial neural networks (all neurons of each layer are connected to the other layer) to control synthetic jets located on the walls of a cylinder. The agent interacts with an environment that uses FEnics to simulate the flow and actions taken by the agent. As rotation is a drag reduction method in AFC of fluid mechanics\cite{MITTAL, Nobari}, later on, we added rotation to the cylinder and reported the DRL limits and behavior in such situations. and its behavior. Computing the optimum number of jets and their locations, finding the number and the location of sensors needed for the agent to have an effective observation of the flow field, adding rotation to the cylinder, and observing the DRL’s effectiveness in lowering the drag in this situation, its behavior, and limits, suggesting effective control parameters for the jet configurations with and without the cylinder rotation, and comparison of the best-performing configuration with a researched configuration will be examined. Such results have not been investigated by other researchers yet.

\section{Methodology}
In this section, we will summarize the methodology for performing the numerical simulations and details of the DRL algorithm.

\subsection{\label{sec:level1}Simulation environment}
The simulation's geometry is based on the 2D test case developed by Schäfer et al.\cite{Schäfer}. It consists of a cylinder with a non-dimensional diameter of D = 1, situated in a box with a total non-dimensional length of L = 22 (along the X-axis) and a height of H = 4.1 (along the Y-axis). The coordinate system's origin is located at the cylinder's center, which is 0.05 off-axis in the y direction. The geometry is shown in figure~\label{Geo2}\ref{Geo2}. The inflow profile formula follows a parabolic function which is expressed as \ref{u1}:

\begin{equation}\label{u1}
	U{(y)}={6(H⁄2-y)(H⁄2+y)}/H^2
\end{equation}

The no-slip boundary condition is applied to the top and bottom walls, as well as the solid walls of the cylinder. In the case of rotation, the speeds are applied over the boundary of the cylinder's wall. On the right side of the domain, an outflow boundary condition is enforced. The Reynolds number is calculated using the average velocity magnitude and the diameter of the cylinder $(Re = \bar{U}D/\nu )$, with $\nu$ being the kinematic viscosity, $\bar{U} = 2U(0)/3=1$ and $Re = 100$. To perform computations, an unstructured mesh has been created using Gmsh \cite{Geuzaine}. This mesh is made up of $9252$ triangular elements and is refined around the cylinder. The simulation utilizes a consistent non-dimensional time interval of $dt = 5\times10^{-4}$. The instantaneous drag force on the cylinder is computed as follows:

\begin{equation}
	F_D=\int\limits_c (\sigma.n).e_x \ dS \
\end{equation}
This equation involves the Cauchy stress tensor (represented by $\sigma$), the unit vector normal to the outer cylinder surface (represented by $n$), and the vector $e_x$, which has a value of (1,0). From this equation, the drag force can be normalized into the drag coefficient:
\begin{equation}
	C_D=[\frac{F_D}{1/2\rho\bar{U}_2 D}]
\end{equation}

And the lift is calculated as follows:

\begin{equation}\label{fl}
	F_L=\int\limits_c (\sigma.n).e_y \ dS \
\end{equation}

\begin{equation}\label{cl}
	C_L=[\frac{F_L}{1/2\rho\bar{U}_2 D}]
\end{equation}

\begin{figure}
	\centering
	\includegraphics[width=1\textwidth]{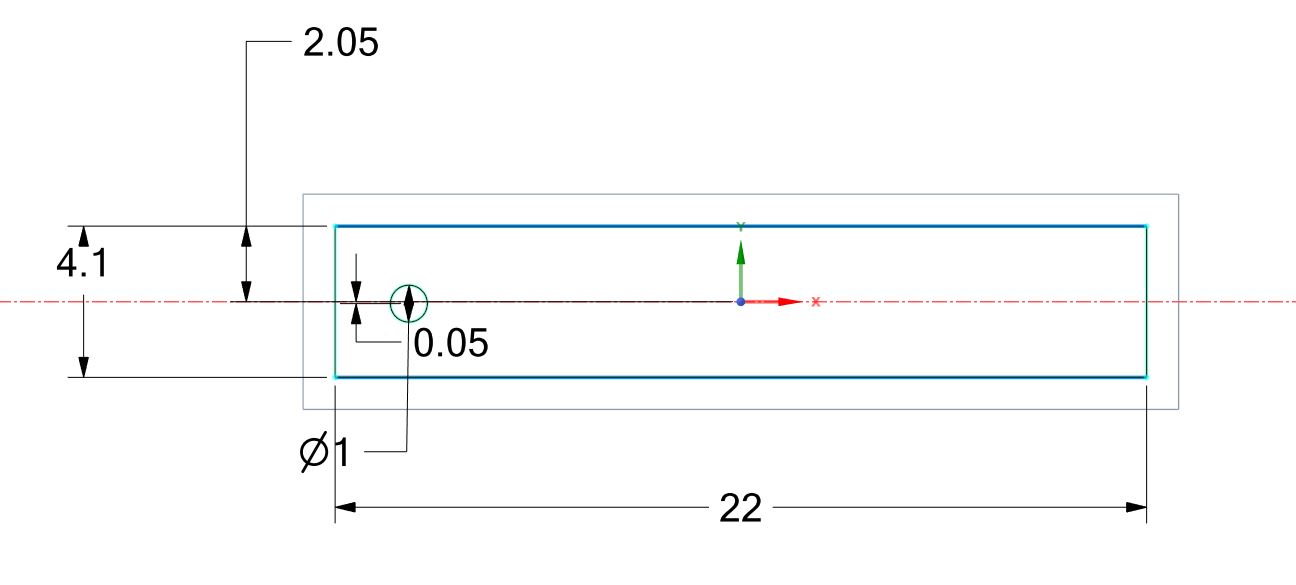}
	\caption{Schematics of the computational domain. \label{Geo2}}
\end{figure}

The IPCS method, developed by Goda \cite{Goda}, is used to solve the Navier-Stokes equations in a segregated manner, and the non-linear term is treated explicitly in this method. The FEniCS framework is used to implement the finite-element method for spatial discretization \cite{Logg}. A maximum of six jets (numbered 0 to 5) normal to the cylinder wall is implemented on the surface of the cylinder, at angles $\theta_0$ to $\theta_5$ relative to the flow direction. The control of these jets is achieved by adjusting their non-dimensional mass flow rates, denoted by $Q_i$, where i ranges from 0 to 5. These flow rates are set by a parabolic velocity profile that reaches zero at the edges of the jets. Additionally, the width of them is $10^{\circ}$.

Furthermore, the control scheme is arranged to ensure that the overall mass flow rate introduced by the synthetic jets is zero, meaning that the sum of all the mass flow rates ($Q_0$ to $Q_i$) is equivalent to zero. This particular condition for the synthetic jets is preferred because it is more ideal than a situation where the mass flow rate is either added or subtracted from the flow. In total, we ran 34 simulations (excluding the validation and test simulations). To summarize the results of these simulations, we grouped them based on the jet positions as shown in the table~\ref{Configurationk}.

\begin{table}
	\centering
	\caption{\label{Configurationk} Details of different simulated configurations.}
	\includegraphics[width=1\textwidth]{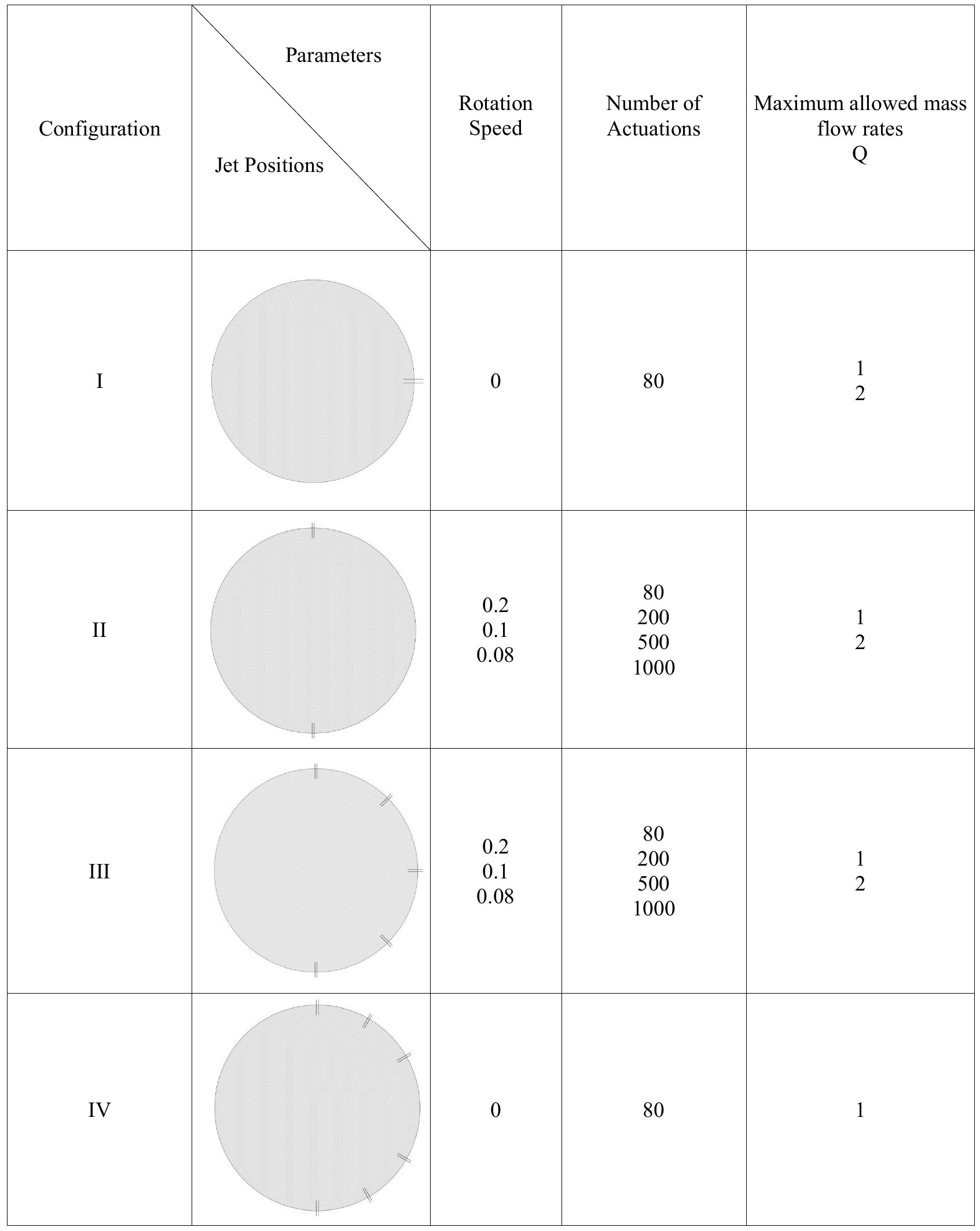}
\end{table}

\subsubsection{\label{sec:level2}The network and reinforcement learning framework}
Here Deep reinforcement learning (DRL) agent sees the simulation as another environment to interact with through three channels namely: observation, action, and reward. Here, the reward is the time-averaged drag coefficient calculated by simulation and provided by the environment then punished by the averaged lift coefficient.
The DRL uses this limited information to train a deep neural network. The network discovers a closed-loop control strategy by observing each time step and deciding on the action, which in this case is the flow rate of the jets. The goal is to maximize the reward, which is the suppression of drag force.
As stated before, the DRL agent uses the proximal policy optimization \cite{Schulman} method. The proximal policy optimization is a type of reinforcement learning algorithm that falls under the category of policy gradient methods. There are several reasons why this method was chosen. For one, it's mathematically less complex, and as a result, it is faster than other methods like the trust region policy techniques \cite{Schulman}. Additionally, it requires less hyperparameter tuning and is easier to set up. It's also well-suited for problems that involve continuous control, which sets it apart from the deep Q-learning \cite{Mnih} and its variations \cite{Gu}.

The PPO method follows an episode-based approach, where it learns from actively controlling for a limited period of time before analyzing the results and continuing with a new episode. In our case, we initially run the simulation without active control until an unsteady wake is developed at approximately 5 seconds. Then this state is saved and serves as the starting point for each subsequent learning episode. The reward function, represented as $r$, is calculated as follows:
\begin{equation}
	r=-⟨C_D ⟩_t-0.2|⟨C_L ⟩_t |
\end{equation}
The goal of the agent is to reduce drag while mitigating the lift fluctuation. $C_D$ and $C_L$ are the drag and lift coefficients, respectively. Here $<.>_t$ indicates an average over a typical vortex shedding period.

There are several benefits in using this particular reward function rather than just the instantaneous drag coefficient. Firstly, by averaging values over one vortex shedding cycle, the reward function becomes less variable, which improves learning speed and stability. Secondly, a penalization term based on the lift coefficient is necessary to prevent the network from cheating. Without this penalization, the ANN can modify the flow configuration to achieve a larger drag reduction, but at the expense of a large induced lift, which can be harmful in practical applications.
We also used Eq.~\ref{smooth} to control instantaneous actuation. This is used so that the agent won't set a high minus flow rate (suction) and instantly after it, a high plus flow rate (blow) as this case is not possible in reality.
\begin{equation} \label{smooth}
	Smooth \: Control= \frac{Number \: of \: actuations}{dt}\times{\frac{0.1\times{dt}}{Number \: of \: actuations}}
\end{equation}

The artificial neural network (ANN) used in this study comprises of two dense layers containing 512 fully connected neurons each, in addition to the necessary layers for data acquisition from the probes and data generation for jets. The configuration of this network was determined through trial and error. Larger networks were found to be less effective due to their increased difficulty in training and longer run times Additionally, the results of the validation showed that utilizing larger networks does not yield any improvements in terms of performance. A learning rate of 1e-3, a sub-sampling fraction of 0.2, and a batch size of 20 are used. Our code is based on the original paper of Rebault et al.\cite{Rebault19} with some tweaks and changes, as we started our work right after their publication.

As for the reported data after training, in most cases, the simulation of flow needed more than 8000 steps (each step is $5\times10^{-4}$), for the controlled flow to stabilize. Also since some cases needed more than 1800 epochs to find a suitable strategy, we ran all the tests with at least 2100 epochs. These points were not reported before and were ignored in previous publications although they are very important.


\section{Validation/verification}
In this section, we discuss the validations done for the flow solver, the time-step independence, the mesh convergence, the sensors' numbers and positions, and finally the ANN network's depth. Validation of the flow solver was performed by observing the drag coefficient and the Strouhal number $St = fD/\bar{U}$, where f is the vortex shedding frequency calculated by performing FFT on lift coefficients. As it can be seen in table~\ref{validation} our $C_{D,max}$ and Strouhal number were in agreement with the data of Tang et al.\cite{Hongwei} and Schäfer et al.\cite{Schäfer} ($0.17\% $ difference in St number. using the main mesh as discussed below). Also, a comparison with the control model of Rebault et al.\cite{Rebault19} was done, and they were in agreement with each other (considering the two decimal number that was reported).

\begin{table}[h]
	\centering
	\caption{Details of the flow solver validation. \label{validation}}
	\includegraphics[width=1\textwidth]{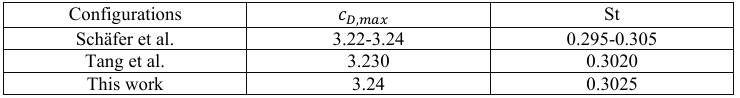}
\end{table}

In this work, a non-dimensional constant numerical time-step of $dt = 5\times10^{-4}$ is used. To confirm that this time-step is small enough, we compared it to $dt = 1\times10^{-4}$ and $dt = 7\times10^{-4}$. As shown in table~\ref{dt} the maximum difference is at $0.47\% $ which is acceptable if we factor in the calculation time difference.

\begin{table}
	\centering
	\caption{Results of the time-step independence study. \label{dt}}
	\includegraphics[width=1\textwidth]{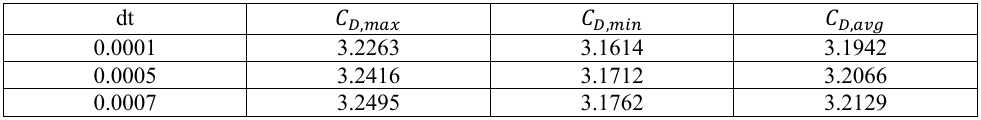}
\end{table}

\begin{figure}[h]
	\centering
	\includegraphics[width=1\textwidth]{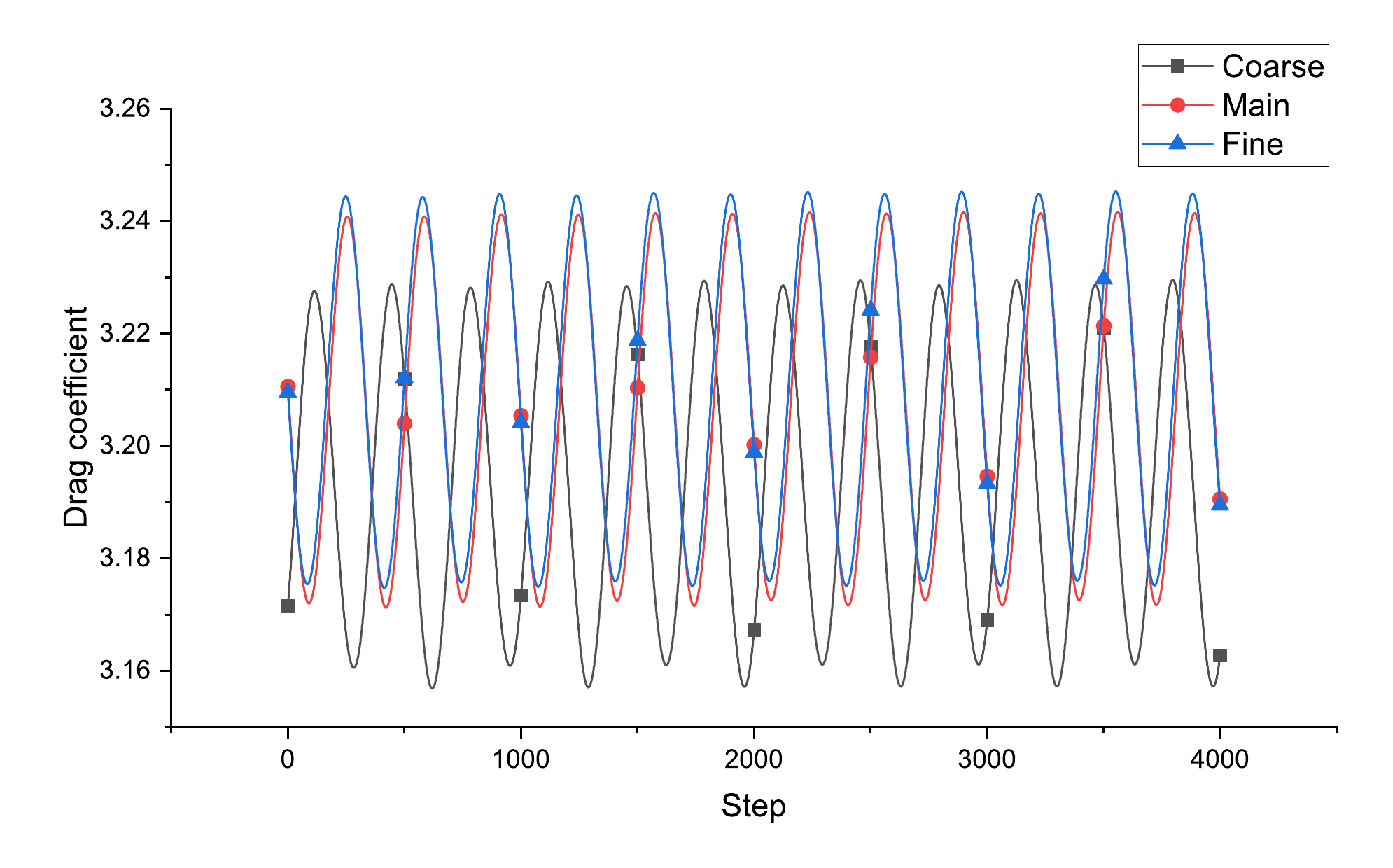}
	\caption{variation of the drag coefficient with respect to the steps obtained from three meshes.}
	\label{chart:meshdraw}
\end{figure}

The mesh-independence study for this benchmark was done with the meshes of three different resolutions, as can be seen in table~\ref{MESH}. The resolution of the mesh which we used in this paper, is fine enough for the simulation and fast enough for a CPU with 16 threads. Figure~\ref{chart:meshdraw} compares the line chart of the drag coefficients with the difference between the main and fine being only $0.11\%$.
The overall mesh and a snapshot of the mesh near walls used in simulations are shown in figure~\label{MESH2d}\ref{MESH2d}.

\begin{table*}
	\centering
	\caption{Report of the mesh independence study.}
	\label{MESH}
	\includegraphics[width=1\textwidth]{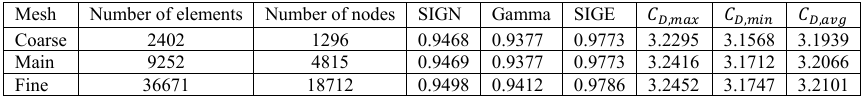}
\end{table*}

\begin{figure}
	\centering
	\includegraphics[width=1\textwidth]{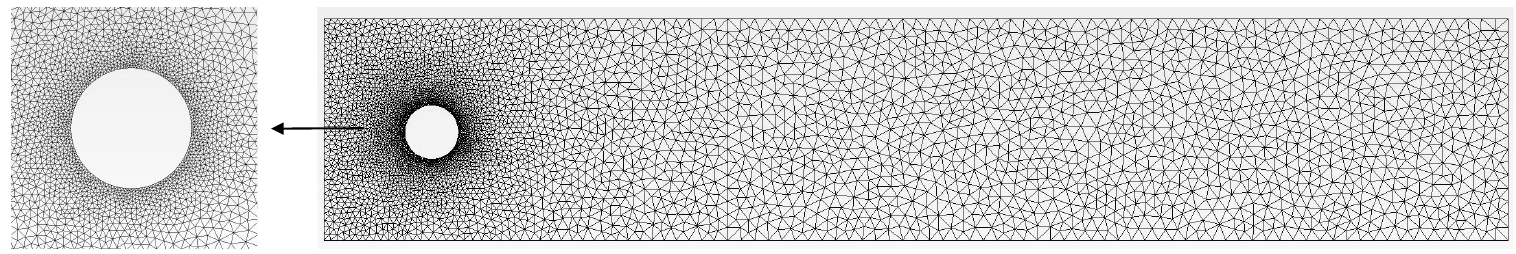}
	\caption{Two-dimensional unstructured mesh of the computation domain of the flow over a circular cylinder and a close snapshot near walls. }
	\label{MESH2d}
\end{figure}

For the main work, we used a total number of 151 sensors and tested up to 600 sensors in different locations in the channel. We performed these tests on configuration III with 200 actuations per episode, no spin, and the maximum allowed flow rate of each jet actuation set to 1. The numbers in the table were extracted after the stabilization of the controlled flow (except the averaged $C_D$), (figure~\label{ns2}\ref{ns2}) and we ran each test multiple times with 2100 episodes and higher. As it can be seen in table~\label{Ns1}\ref{Ns1} more sensors in other words more data for the agent does not always lead to better performance. We believe that this fact is because when we add data of the far reach of the cylinder, our network tries to over-control those locations. So, the performance near the cylinder gets worse.

We should also note that the most efficient number and position of sensors are different for each configuration with different parameters. One such example is when we use 123 sensors from 0.075 until 0.25 (the same configuration as the tests presented here). In this case, $C_{D,max}$ goes up from 2.2047 to 2.2240. This is an increase of $0.87\%$ compared to the 151 sensors, which in some cases may be acceptable. But it is not acceptable for configuration III with 1000 actuations per episode/epoch, 0.2 spin, and maximum allowed flow rate of each jet actuation set to 2, because in this case our $C_{D,max}$ increases from 1.6288 to 1.8235 which is an $11.95\% $ difference compared to the main sensor number and position. Hence, we opted to use the main sensor configuration as it provides sufficient data for the agent in almost all cases.

\begin{figure}
	\centering
	\includegraphics[width=1\textwidth]{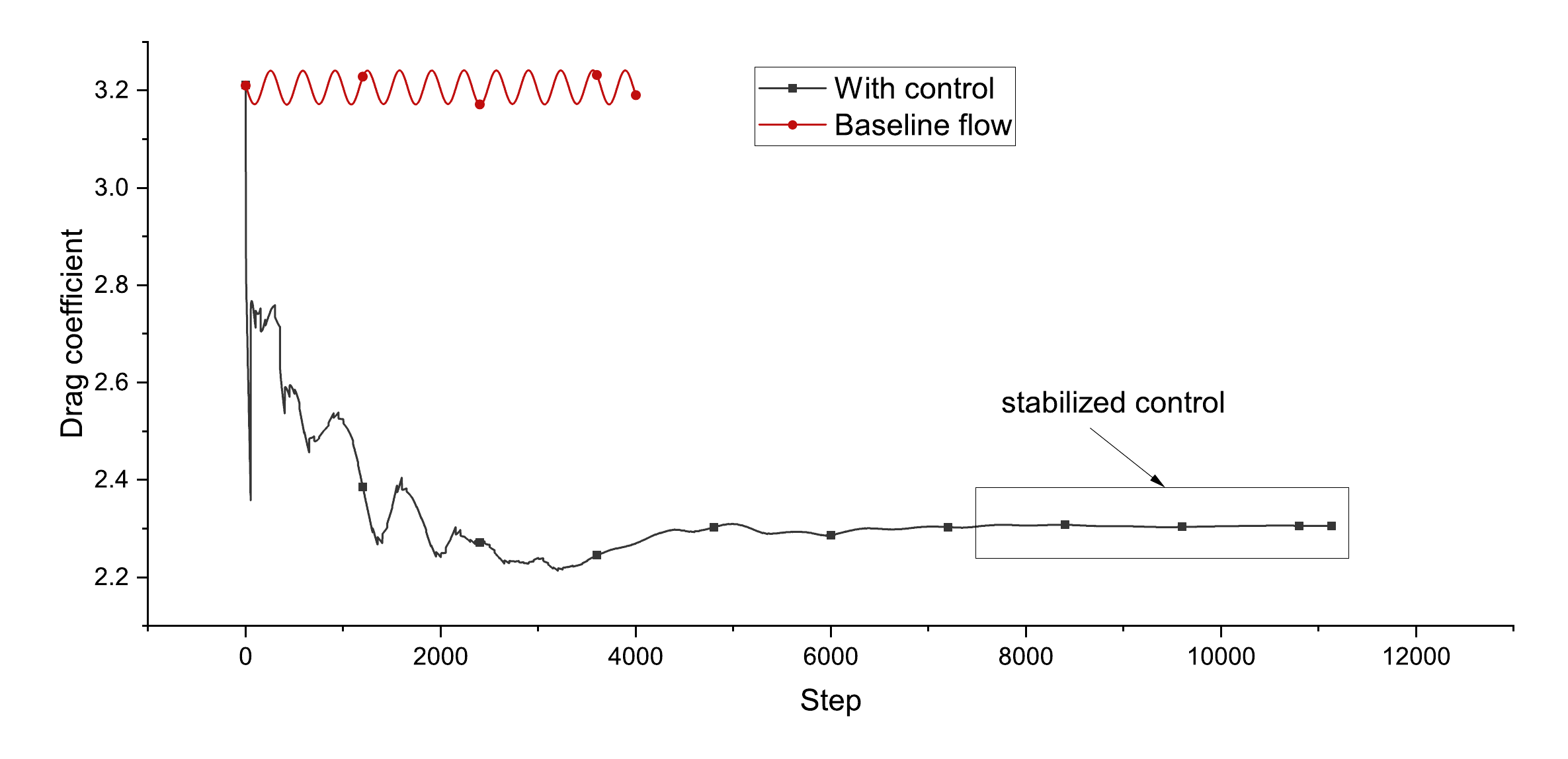}
	\caption{The drag coefficient for the controlled flow with respect to the number of steps. The numbers were extracted from the region inside the rectangle.\label{ns2}}
\end{figure}

\begin{table}
	\centering
	\caption{Comparing the minimum, maximum, and average drag coefficients of the different sensor numbers and the area occupied by them for configuration III with the no-spin, the maximum flow rate of 1 and 200 actuations.\label{Ns1}}
	\includegraphics[width=1\textwidth]{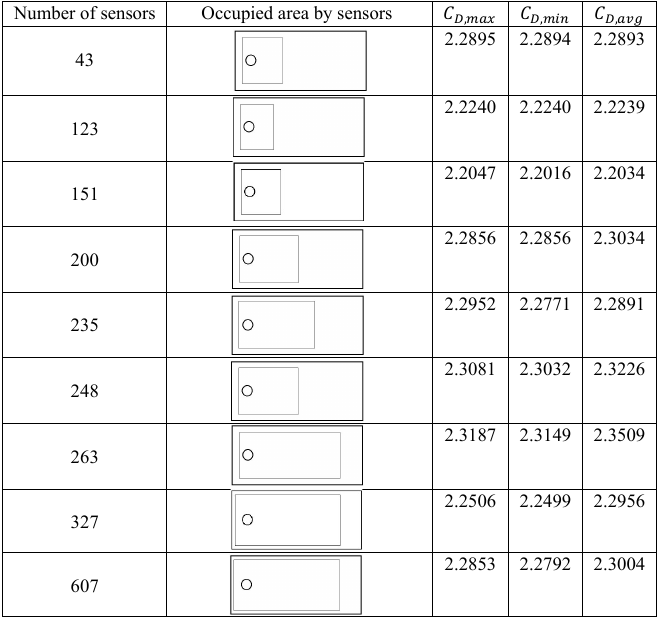}
\end{table}

Test for the network depth needed for this use case was done with four 512 fully connected networks instead of the two that we are using in our work. As it can be seen in figure~\label{net}\ref{net}, using a deeper network does not bring any benefit, and makes the run-time more than twice longer compared to the two 512 fully connected networks. Also, deeper network with less total neurons (four 128 networks) were compared to a shallower network with a higher amount of neurons (two 512 networks). It was observed that the deeper network with lower total neurons (512) was 0.3\% faster, had an 18.1\% lower cpu-time (7.73 compared to 6.55 minutes), but performed 5.11\% worse. To make sure that the network was trained long enough and properly, we ran this part with up to 4200 epochs and with different learning rates and entropy regularization numbers. The learning rate and entropy regularization are two of the most important parameters during training a network. So, these two parameters were tested with different numbers to check for possible hyperparameter balancing. However, the finest results are shown in figure~\label{net}\ref{net}.

\begin{figure}[h]
	\centering
	\includegraphics[width=1\textwidth]{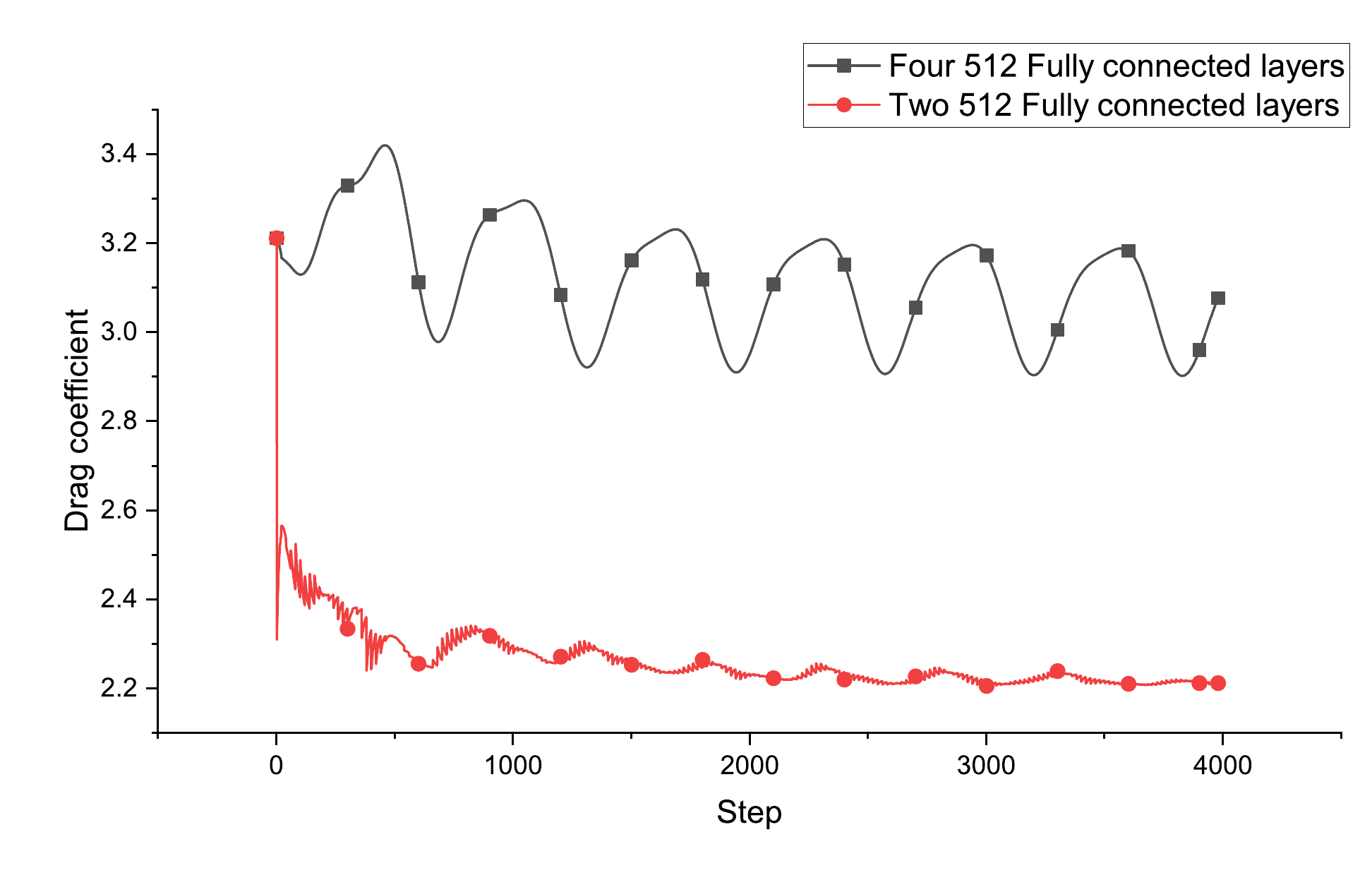}
	\caption{Drag coefficient values of two networks with respect to the number of steps. \label{net}}
\end{figure}


\section{Results}
Here, we will find the best-performing configuration regarding drag reduction among the tested ones. First, we will test different parameters to find the best control parameters for the most efficient case, and then, the rotary motion will be added to the cylinder. Finally, we will compare the most efficient jet configuration with configuration II which was mostly used in other works, and test the results with different control parameters as before to find the best-performing ones.

\subsection{\label{sec:level1}The no-spin case}
First, in order to find the optimum performing configuration regarding the number and location of the jets, we used configurations I through IV with the default parameters of the no-spin, the maximum allowed flow rate of each actuation set to 1, and the number of actuation per episode set to 80.
Configuration I has one jet placed at ${0^\circ}$, the case II has two jets located at ${90^\circ}$, ${270^\circ}$, the case III has five jets inserted at $0^{\circ}$, $45^{\circ}$, ${90^\circ}$, ${270^\circ}$, ${315^\circ}$ and configuration IV has six jets placed at ${30^\circ}$, ${60^\circ}$, ${90^\circ}$, ${270^\circ}$, ${300^\circ}$, ${330^\circ}$, all jets are numbered 0 to 5.

Table~\ref{bestconfiguration} shows the start, minimum, maximum, and average drag coefficients of the mentioned configurations. The minimum and maximum values were extracted after the stabilization of the controlled flow, but the average values are on the whole episode. The start drag coefficient refers to the $C_D$ at the start of an episode (which is the center point of $C_D$ when there is no control). As can be seen in this table, the most efficient case is the third configuration. with a drag reduction of $35.06\%$ (from $C_{Dmax}$ without control). So, we kept this case for our next studies.

\begin{figure}
	\centering
	\includegraphics[width=1\textwidth]{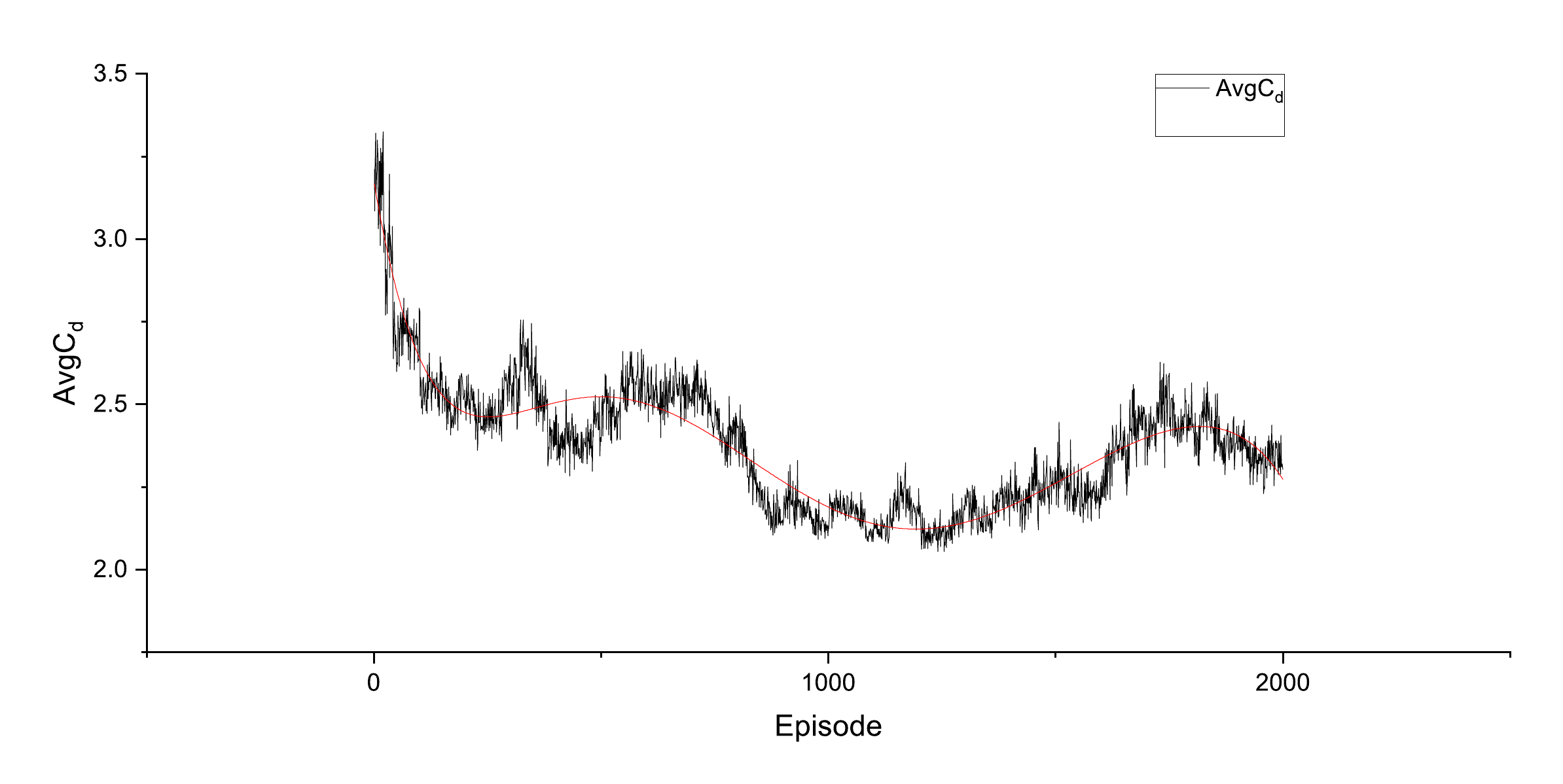}
	\caption{\label{Cd5jet} The average drag coefficient during 2000 episodes of the learning process.}
\end{figure}

\begin{figure}
	\centering
	\includegraphics[width=1\textwidth]{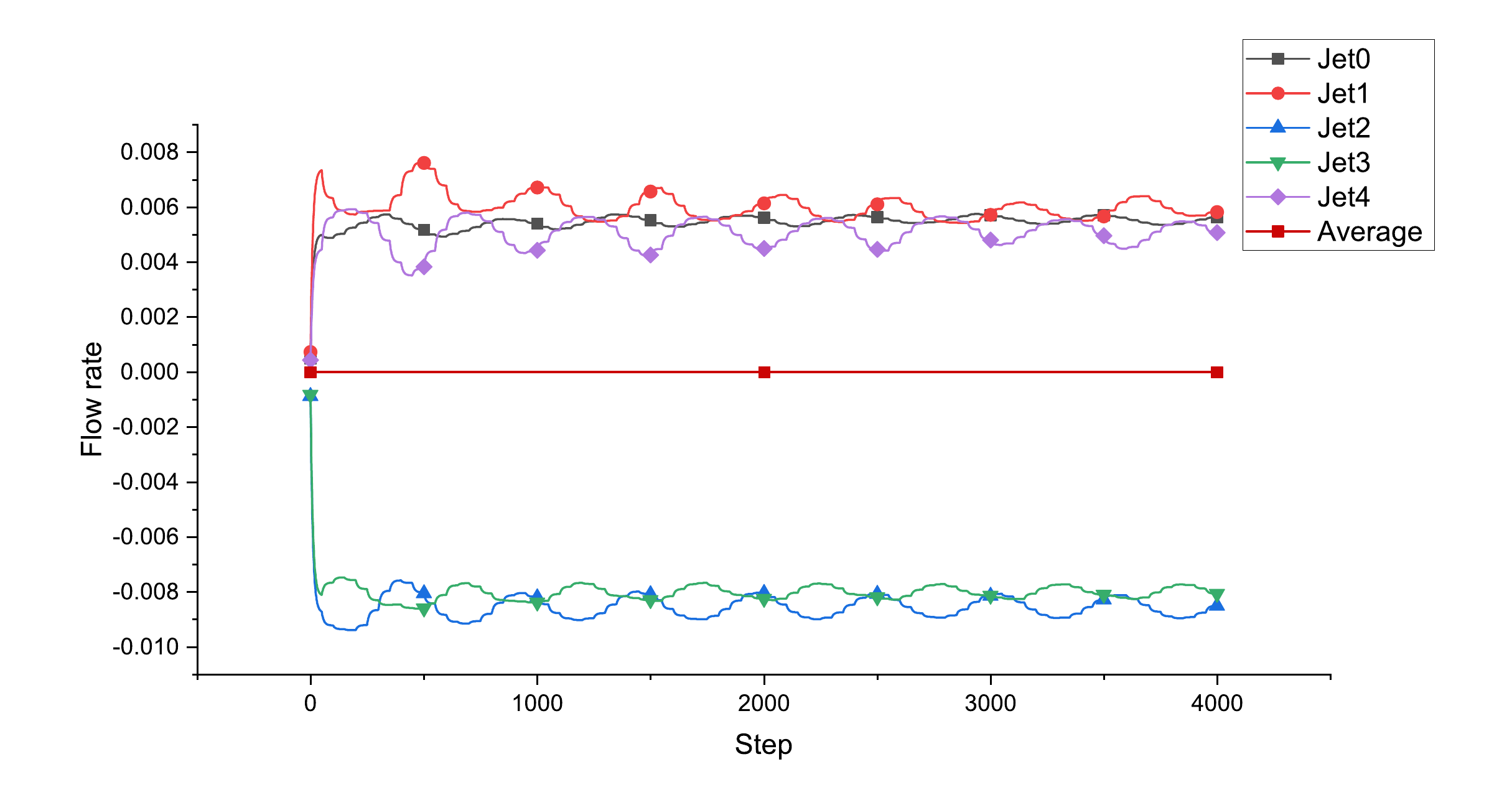}
	\caption{\label{flowrateo} The flow rate of each jet during control and their average value, which sits at zero.}
\end{figure}

Figure~\ref{Cd5jet} shows the variation of $C_D$ during the learning process. Each point on this chart belongs to an epoch. As can be seen in this chart, we need around 1300 epochs at the very least for the agent to find the best controlling strategy, and because of this, we ran each case with 2100 epochs or higher. The agent needed 817, 1106, 1241, and 1237 epochs for configurations I to IV, respectively in order to find the best controlling strategy. Figure~\ref{vel5}\label{vel5} illustrates the pressure and vorticity field of this configuration at the start and the end, after flow stabilization. As can be seen in these figures, the wake is thinner in the case with active control, which results in a lower mean pressure difference around the cylinder and lower drag force. Another point that can be seen from the vorticity field is that separation is delayed due to active control. The position of the jets is obvious in close snapshots. The jet flow rates for this configuration are shown in figure~\ref{flowrateo}, which after around 1500 steps stabilize.

\begin{table}
	\centering
	\caption{\label{bestconfiguration} The mean value of the drag coefficient at the start of the control, the maximum and the minimum CD when the control is stabilized, and the average CD on the whole controlled episode.}
	\includegraphics[width=1\textwidth]{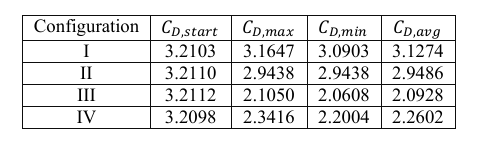}
\end{table}

\begin{figure}
	\centering
	\includegraphics[width=1\textwidth]{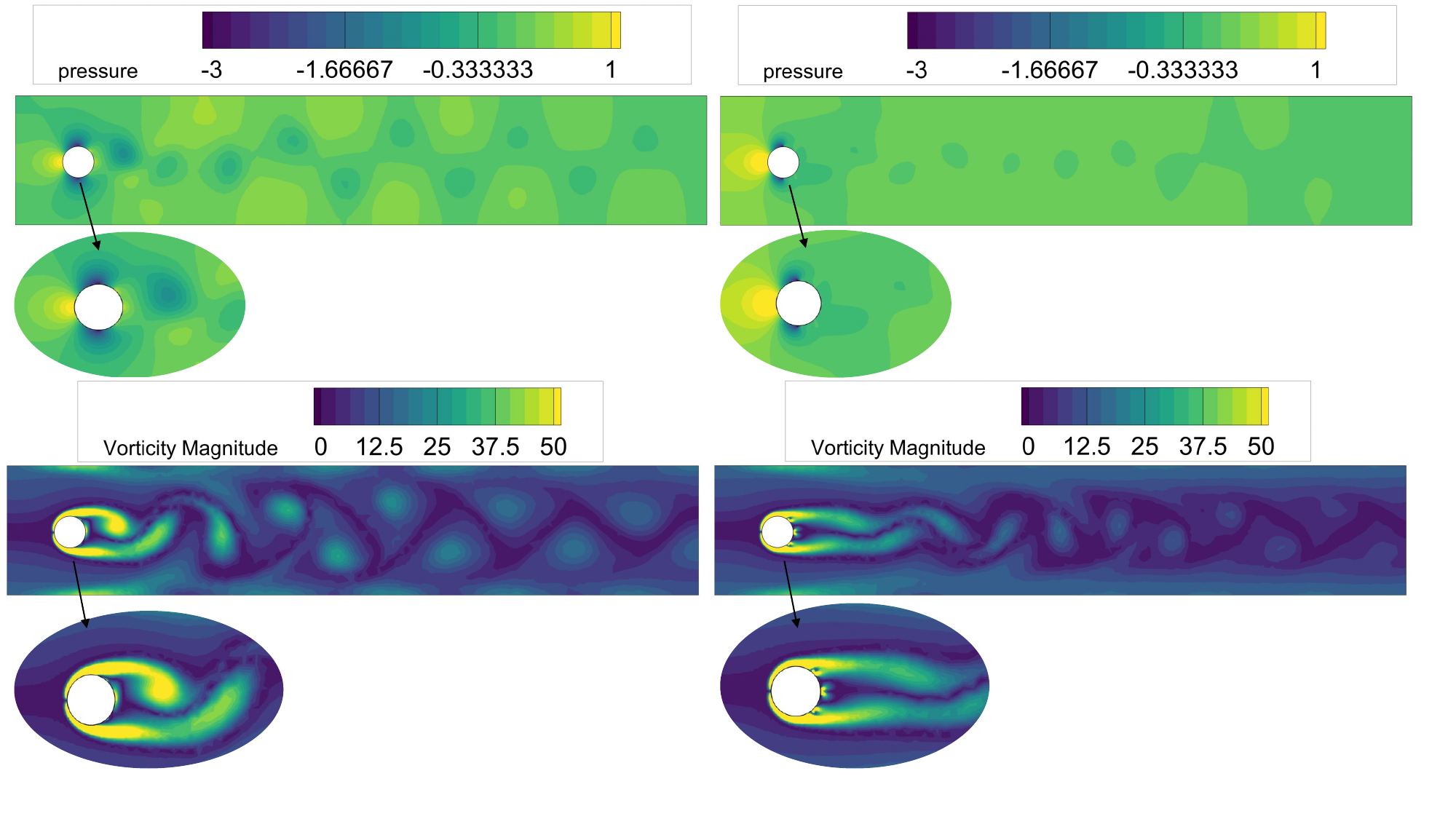}
	\caption{Pressure and vorticity contours for configuration III at the start (left) and the end of control (right). \label{vel5}}
\end{figure}

Now that we have found the optimum jet configuration, it will be tested with different parameters in order to find the most efficient performing system with the purpose of lowering the drag force. The spin, the number of actuation per episode, and the maximum allowed flow rate of each actuation will be compared with the same parameters in configuration II.

\begin{figure}
	\centering
	\includegraphics[width=1\textwidth]{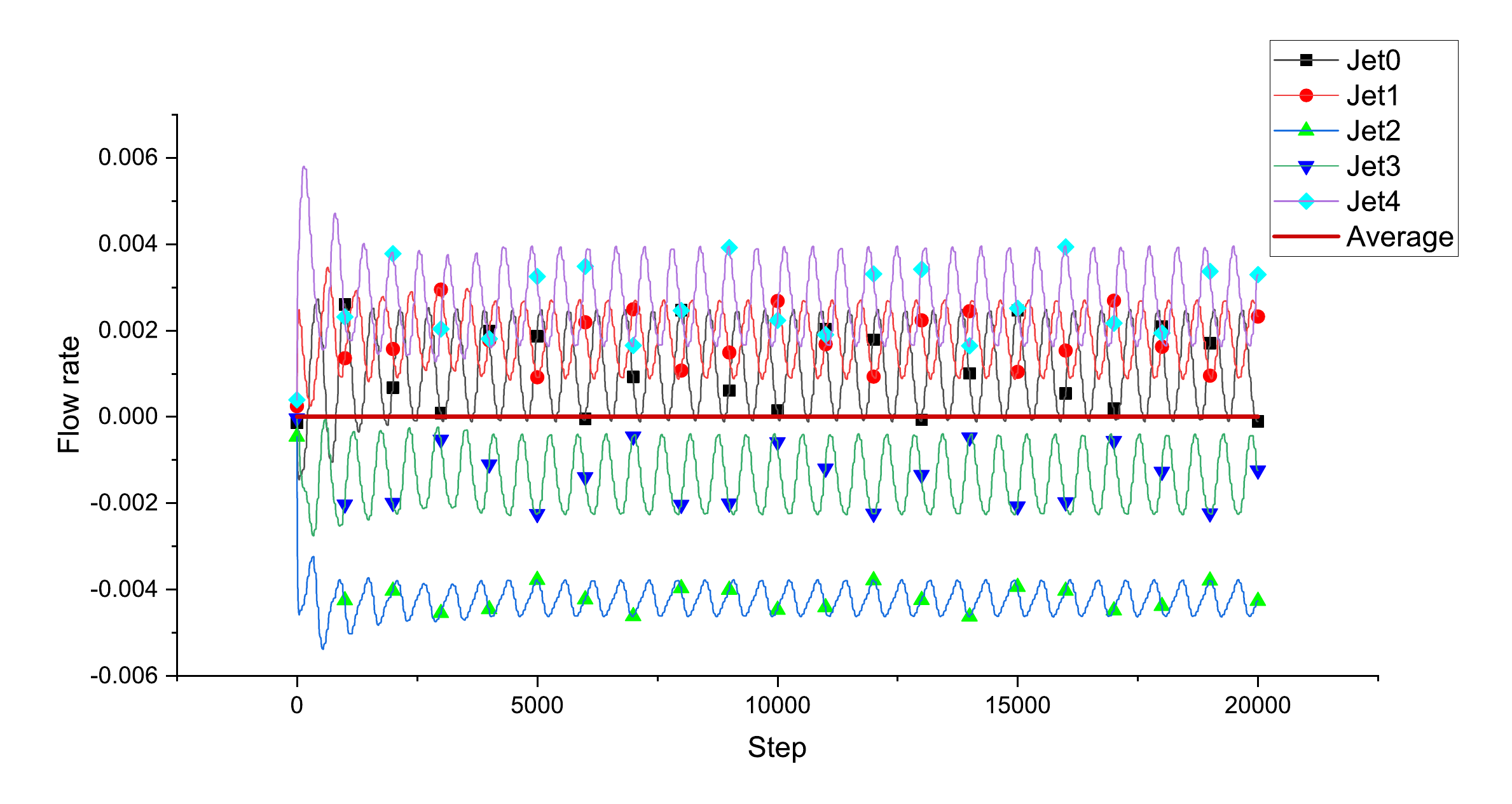}
	\caption{\label{Qnbm} The flow rate of each jet during the control process.}
\end{figure}

In the first test, we doubled the maximum allowed flow rate of each actuation, and performance got worse. Its maximum, minimum, and average $C_D$ arises from 2.1050 to 2.8053, from 2.0608 to 2.6171, and from 2.0928 to 2.6980 respectively, which translates to an uplift of $33.27\% $, $26.99\% $ and $28.92\% $ in $C_D$. The variance of $C_D$ when the control is stabilized in this case is also higher at $1.42\times10^{-3}$ compared to $3.26\times10^{-5}$. As it can be seen from the jet flow rates shown in figure~\ref{Qnbm}, compared to the figure~\ref{flowrateo}, it seems that the agent struggles to stabilize the flow, which can be the cause of this trend. In this case, the agent needed only 40 epochs in order to find the optimum model.

Next, we tried to change the number of actuation per episode from 80 to 200,500,1000 in order to see its effects as compared to doubling the maximum allowed flow rate of each actuation. The agent has the possibility of using larger flow rates like the case before, just by using different means. First, we increased the number of actuation from 80 to 200 and observed that our system performed $21.41\%$ better than when we used a higher allowed flow rate for each actuation. But it performed $4.74\% $ worse compared to the test with 80 actuations. The computed $C_D$ in this case were 2.2047, 2.2016, and 2.2034 for minimum, maximum, and average values, respectively. The $C_D$ variance in this case was $9.24\times10^{-7}$ and the agent found the best model at episode number 1805.

Further increasing the number of actuation to 500 results in maximum, minimum, and average $C_D$ of 2.2804, 2.2544, 2.2626 respectively, which are worse than when using 80 and 200 as the number of actuation by $8.33\% $ and $3.43\% $ $(C_{Dmax})$. But they are more efficient than the case with $maxQ=2$ by $18.71\% $. In this case, the variance of $C_D$ was $1.43\times10^{-4}$. Also, the agent found an efficient control policy in episode 382. With changing this parameter to 1000, a drop of $1.07\% $ in minimum $C_D$ compared to the case with 500 actuation per episode, from 2.2544 to 2.2302, an increase of $7.05\% $ in maximum $C_D$ from 2.2804 to 2.4411, and an increase of $3.83\% $ in the average $C_D$ was observed. Also, the variance is worse at $5.86\times10^{-3}$ and our agent needed 982 episodes to find the best policy.

Results of the tests showed that in this configuration without spin, allowing the agent to have access to an excessive flow rate is not a good idea. As the finest performing control parameters considering minimum and maximum $C_D$ is with 80 actuation per episode and maximum allowed flow rate of each actuation = 1 ($maxQ=1$). Also, the best-performing parameters in terms of stability were the case with 200 actuation per episode and $maxQ=1$ with a $C_D$ variance of only $9.24\times10^{-7}$ after the flow stabilization. The agent needed the most episodes for 200 actuation per episode and the least for $maxQ=2$ and 80 actuation per episode. We should also mention that the higher number of actuation slows down the simulation speed.

\subsection{\label{sec:level2} The rotating cylinder}
Rotation is a proven way to reduce the drag force. Here, we tested three different rotation speeds of 0.2, 0.1, and 0.08 in conjunction with our DRL network to further reduce the drag coefficient and see if the agent can find an efficient control strategy. Again, we will test them with different numbers of actuation per episode and the maximum allowed flow rate per actuation to find the optimum configuration.

A rotation speed of 0.2 alone reduces $C_{Dstart}$ from 3.2112 to 2.7093, which is a reduction of $15.63\%$, and reduces $C_{Dmax}$, $C_{Dmin}$ and $C_{Davg}$ to 2.7115, 2.7172 and 2.7084. Adding the DRL network (configuration III) with 80 actuation per episode and $maxQ=1$, decreases $C_D$ further to a maximum, minimum, and average of 2.3703, 2.3702, and 2.3702, respectively. Compared to $C_{Dmax}$ after adding rotation, it's a further drop of $12.51\%$, with a variance of only $2.73\times10^{-10}$, which can be said that there are no fluctuations at all. Here, the agent needs 517 episodes to find the finest strategy. We ran configuration II with the same parameters, and our agent couldn't find a control strategy in this jet's configuration that decreased $C_D$ further. When we only doubled our maximum allowed flow rate for both configurations II and III, the network could not find a control strategy even after 2100 epochs.

Increasing the number of actuation per episode to 200, results in the system performing better at 2.3197 for minimum (2.31968), maximum (2.31972), and average (the run time is so long that fluctuations are removed from the averaged $C_D$). The variance is $7.75\times10^{-10}$ which again, is so small that we can ignore the fluctuations after the flow stabilization. The network needed 713 epochs to find the best model. Doubling the maximum allowed flow rate decreases $C_D$ by $10.51\%$ to 2.0758 with a variance of $8.77\times10^{-13}$ which is more stable than before, the best performing strategy was found at the 117th epoch. Again, when we ran configuration II with the same parameters (both $maxQ=1 and 2$), the agent could not find a suitable control strategy.

Using 500 actuations per episode, causes $C_D$ to decrease by $0.87\%$ to 2.2994, compared to the 200 actuations ($maxQ=1$), with a variance of $1.76\times10^{-14}$, Which is also more stable than before. The network found the most efficient strategy at 541 epochs. Performance with Double the $maxQ$ was better by $26.13\%$ at 1.6985 with a variance of $5.07\times10^{-14}$. Training the network needed 391 epochs. As before, configuration II with the same parameters could not find a proper controlling strategy.

In consequence of increasing the number of actuation per episode to 1000 and keeping $maxQ=1$, the $C_D$ increased to 2.3394 but variance was better at $4.54\times10^{-15}$. Furthermore, the agent needed 831 epochs to find the best control strategy. With $maxQ=2$, $C_D$ stabilized at 1.6288, which is the best-performing configuration with a drop in drag coefficient of $39.88\%$ compared to the starting point with spin. $C_D$ variance was also nonexistent at $1.93\times10^{-14}$. The network found the model at 544 epochs, also configuration II could not find a control strategy.

\begin{figure}
	\centering
	\includegraphics[width=1\textwidth]{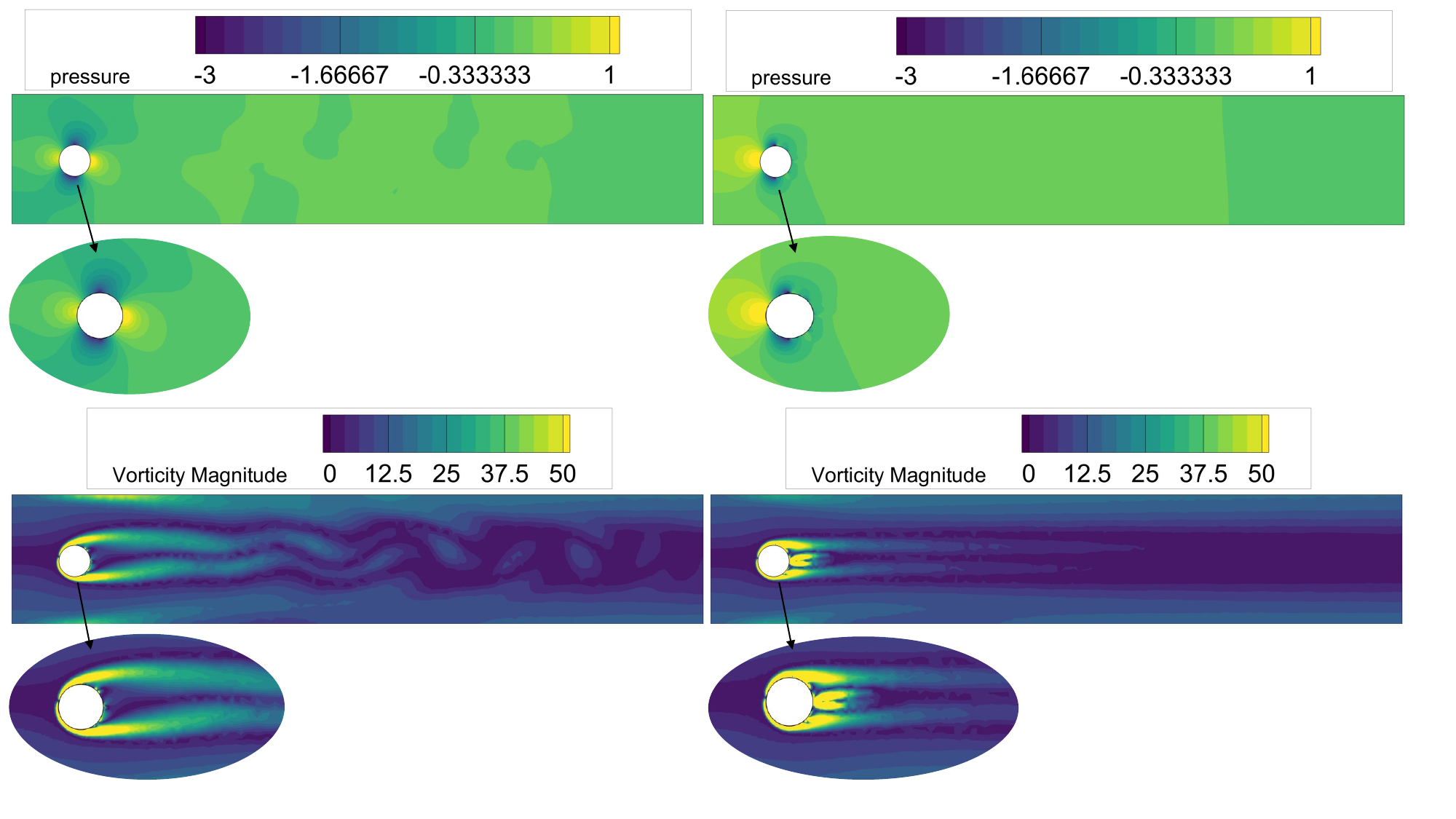}
	\caption{Pressure and vorticity contours at the start (the no-control scenario), the left column; and end of the controlled flow, the right column. \label{spin}}
\end{figure}

In these tests, we found that our agent indeed can control and further lower the drag force exerted on a rotating cylinder. But compared to a fixed cylinder, the agent needs access to a higher amount of flow rate, as it performed best with a maximum allowed flow rate of each actuation = 2 and an increased number of actuation per episode of 1000. In this state, our system managed to lower the $C_D$ from 2.7093 to 1.6288 with a completely stable flow after stabilization. Pressure and vorticity fields of this configuration can be seen in figure~\ref{spin}. The vortex shedding has been completely omitted due to the high momentum of the injected fluid and a high-pressure zone appears behind the cylinder in the wake. The creation of this high-pressure zone in the wake is the main origin of drag reduction.  

As we discussed before, configuration II was unable to lower the drag force when the cylinder had a rotation speed of 0.2. In order to examine the performance of this configuration, and to find its limits when used with a rotating cylinder, we ran further tests with 0.1 and 0.08 spin and compared them with those of configuration III. Indeed configuration II can find a control strategy for rotation speeds of 0.1 and lower. We ran our tests with 80, 200, 500, and 1000 actuation per episode, a $maxQ$ of 1, and compared them with configuration III. Table~\ref{bestconfig} shows the results. $C_D$ variance was $2.44\times10^{-6}$, $2.78\times10^{-7}$, $3.24\times10^{-5}$, $3.87\times10^{-5}$, respectively. The best-performing control parameter for configuration II was 200 actuation per episode with a minimum, maximum, and average $C_D$ of 2.8231, 2.8247, 2.8248. Which is a drop of $12.86\%$ compared to the $C_{Dmax}$ without control. For reference, configuration II without the cylinder rotation could only lower the drag force by $8\%$. Also, it was the most stable, but still, configuration III with 80 actuation per episode was superior with minimum and maximum $C_D$ of 2.0920 and more stable with a variance of only $1.60\times10^{-14}$. The agent needed 681, 342, 530, 1173, and 997 epochs to find the best strategy for each test respective to the data in table~\ref{bestconfig}. Figures \ref{velc}\label{velc} and \ref{volc}\label{volc} compare the pressure and vorticity contours of the finest performing configuration II (200 actuation) with those of configuration III. It is seen that the vortex shedding has been suppressed at the end of the controlled flow in configuration III.    

Doubling the $maxQ$ with 80 actuations per episode has no noticeable benefit nor disadvantage in configuration II as its $C_D$ minimum, maximum, and average are at 2.8671, 2.8721, and 2.8702. But variance gets worse at $3.08\times10^{-6}$. The agent finds the optimum model at 548 epochs, but configuration III with the same parameters worsens noticeably (compared to the 80 actions per episode) with minimum, maximum, and average $C_D$ of 2.5723, 2.6690 and 2.6250 which are $22.96\%$, $27.58\%$ and $25.48\%$ worse, respectively. Variance is also noticeably worse at $4.29\times10^{-4}$. The network takes only 29 epochs to find the best model.

\begin{table}
	\caption{\label{bestconfig}The mean value of the drag coefficient at the start of the control, maximum and minimum CD when the control is stabilized and average CD on the whole controlled episode.}
	\includegraphics[width=1\textwidth]{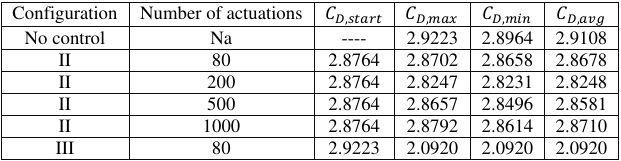}
\end{table}

\begin{figure}
	\centering
	\includegraphics[width=1\textwidth]{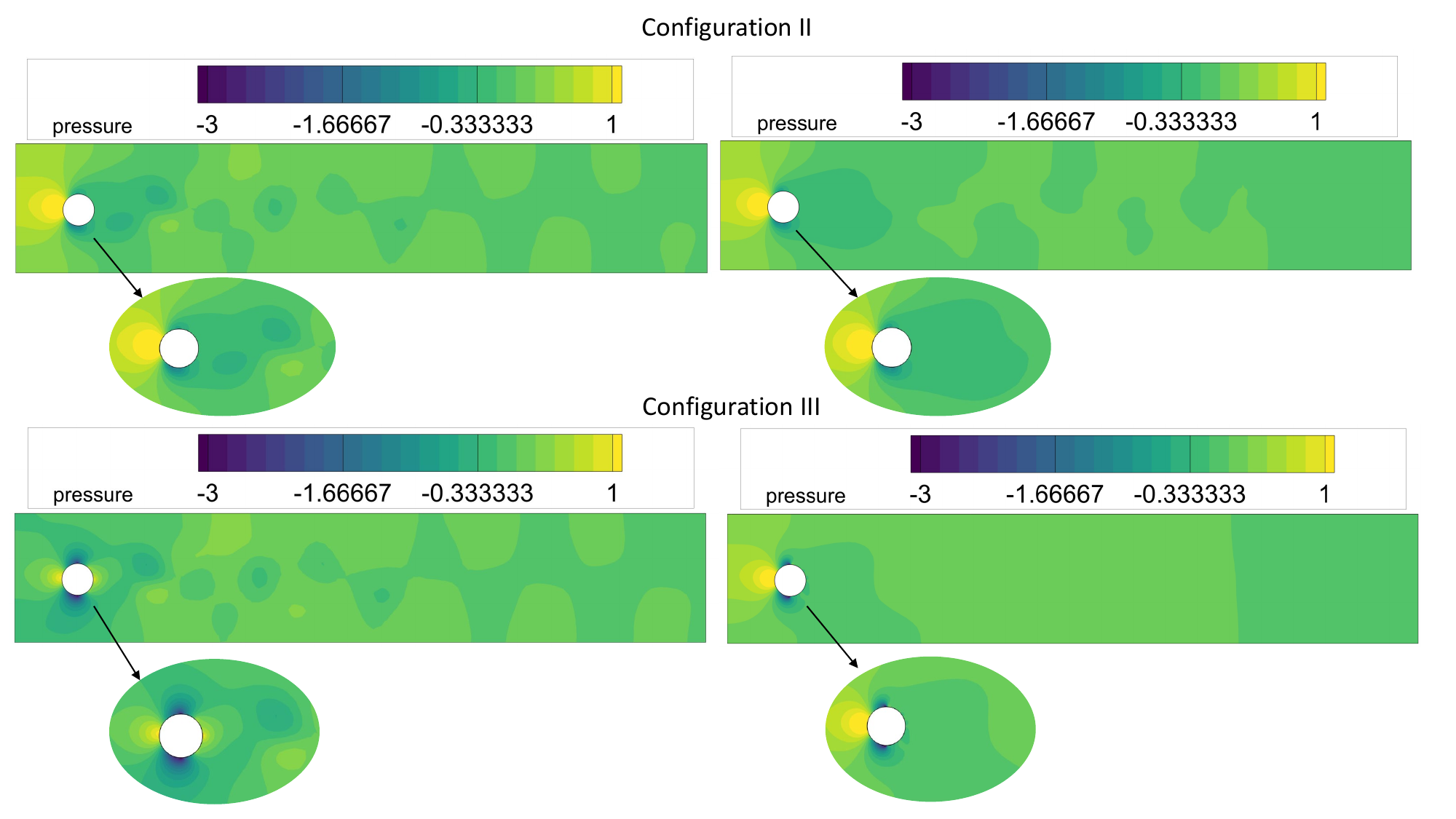}
	\caption{Pressure field at the start (left), and the end of the controlled flow (right) for two configurations. \label{velc}}
\end{figure}

\begin{figure}
	\centering
	\includegraphics[width=1\textwidth]{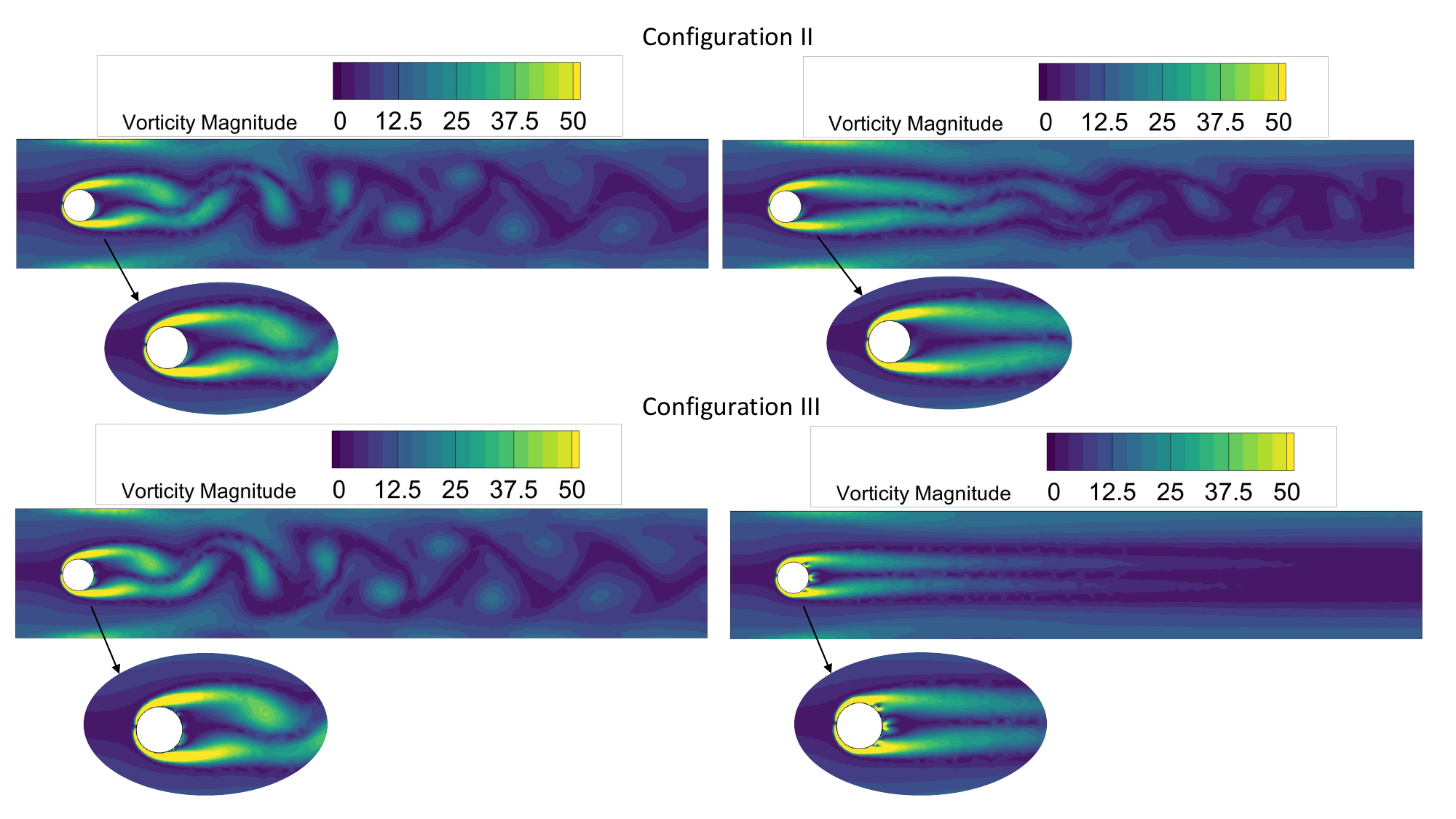}
	\caption{Vorticity field at the start (left), and the end of the controlled flow (right) for two configurations. \label{volc}}
\end{figure}

As the final test, we used 0.08 rotation speed, which without control, $C_D$ was lowered to a maximum, minimum, and average of 2.9738, 2.9448, and 2.9633, respectively. We also tested $maxQ= 1 and 2$ and ran these parameters for configurations II and III with 80 actuations per episode. The start $C_D$ was 2.9353 and configuration II found a strategy to modify it to a minimum of 2.8685, maximum of 2.8854, and average of 2.8776. Doubling the $maxQ$ had no noticeable impact on $C_D$. It lowered the variance from $3.44\times10^{-5}$ to $6.21\times10^{-7}$, but at the cost of higher epochs as this needed 1577 epochs but $maxQ=1$ only needed 22. Configuration III was much better at controlling the drag force as it managed to optimize it to a minimum and maximum of 2.2254 and an average of 2.2254. Also, its $C_D$ variance was much better at $7.71\times10^{-14}$ with the agent needing 1753 episodes to find this strategy. Doubling the $maxQ$ worsens both systems performance (by $15.88\%$) and variance (at $2.94\times10^{-5}$).

\subsection{Real-World Scenario}
\begin{figure}
	\centering
	\includegraphics[width=1\textwidth]{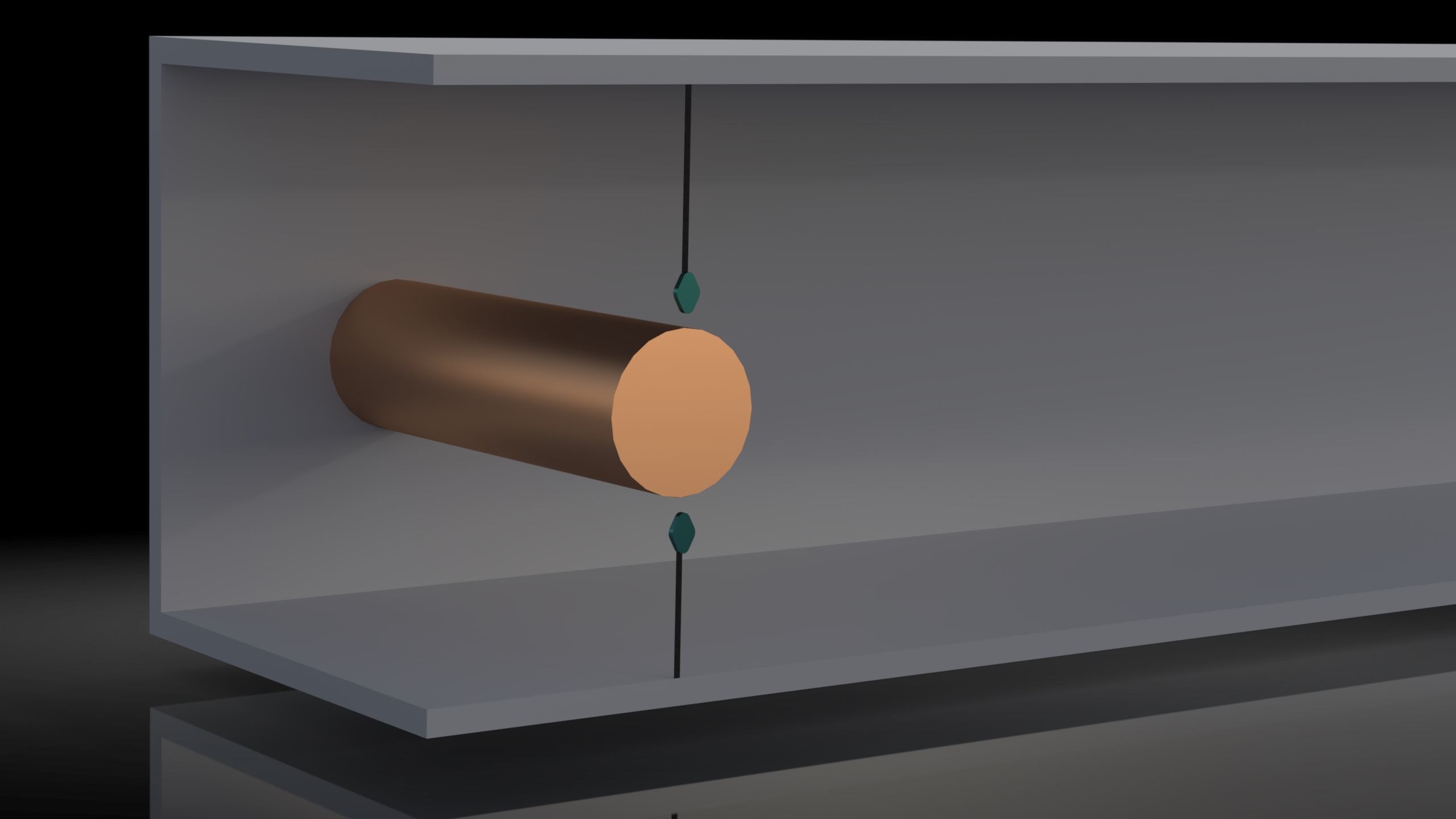}
	\caption{Possible 3D design of the rotating system. \label{3d}}
\end{figure}

In order to show the capability and feasibility of the best-performing configuration for real-world applications, sensors were only placed on the face of the cylinder, and the Reynolds number was increased to 200 (the limit of the accessible hardware). In order to showcase the feasibility of a spinning cylinder a 3D design was created, which is shown in figure \ref{3d}, this is only one of the many possible designs and configurations, and the case in which the cylinder connects to walls. The cylinder can also be disconnected from walls and connected to jet bases for support and rotation. We believe that it is useful for drag reduction in applications like mini-turbines for home use (cases where the cylindrical body of an object is in the wind direction, and rotating it has no adverse effect.).

\begin{figure}
	\centering
	\includegraphics[width=1\textwidth]{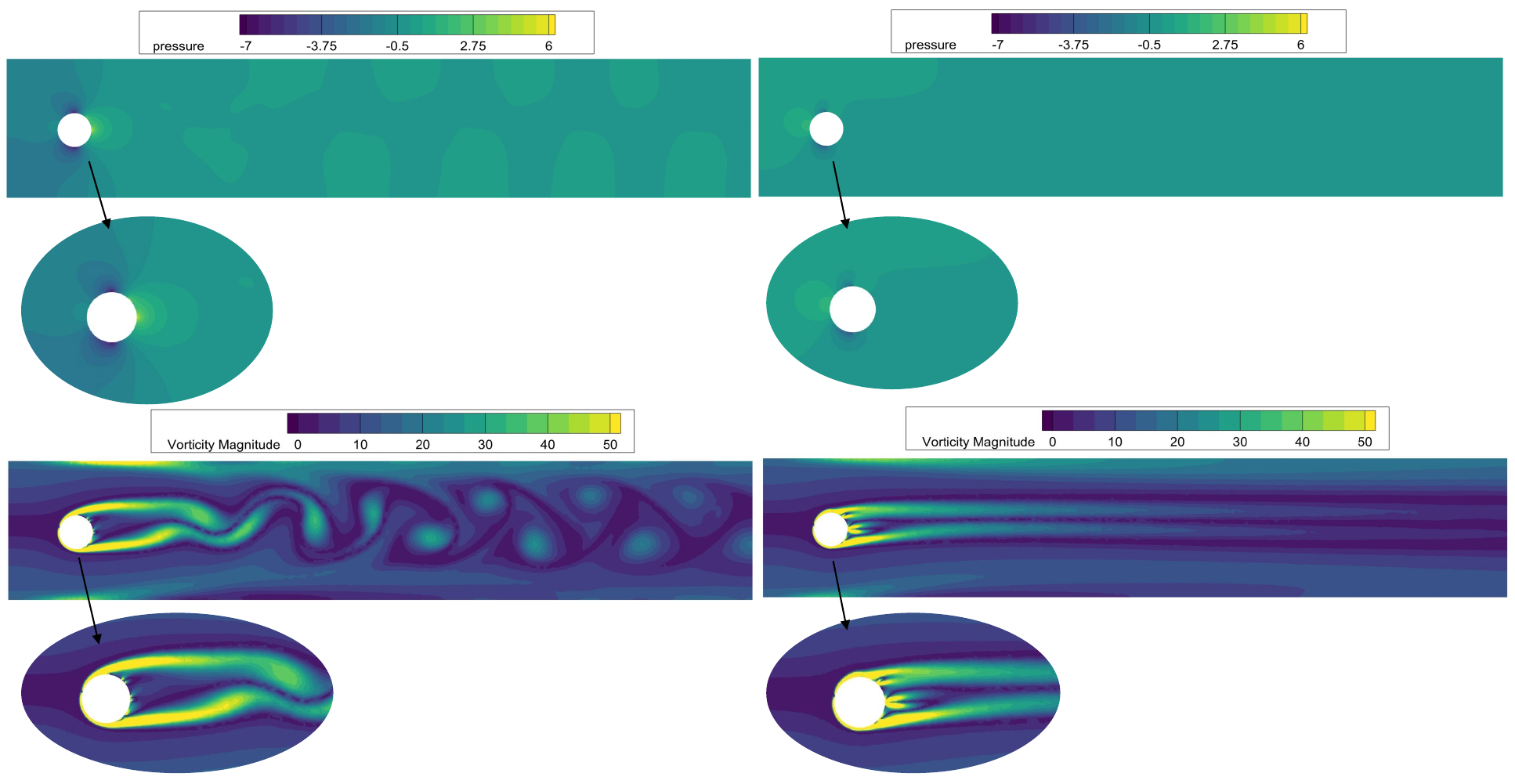}
	\caption{Pressure and vorticity field at the start (left), and the end of the controlled flow (right) for configuration III and Re = 200. \label{200}}
\end{figure}

As a result of limiting the number of sensors to 36 and placing them only on the face of the cylinder every 10 degrees, the system performed worse and the stabilized $C_D$ experienced an increase of 2.72\% from 1.6288 to 1.6731 which is acceptable for real-world applications considering sensors number difference. After doubling the Reynolds number to 200 it was observed that the agent could indeed find a suitable control strategy after 500 epochs. The start drag coefficient was 2.2834 which the DRL agent was able to reduce to a $C_{Dmin}$ of 1.6715, and a $C_{Dmax}$ of 1.6725. Figure \ref{200} shows the vorticity field at the start and the end of the controlled flow.

\begin{figure}
	\centering
	\includegraphics[width=1\textwidth]{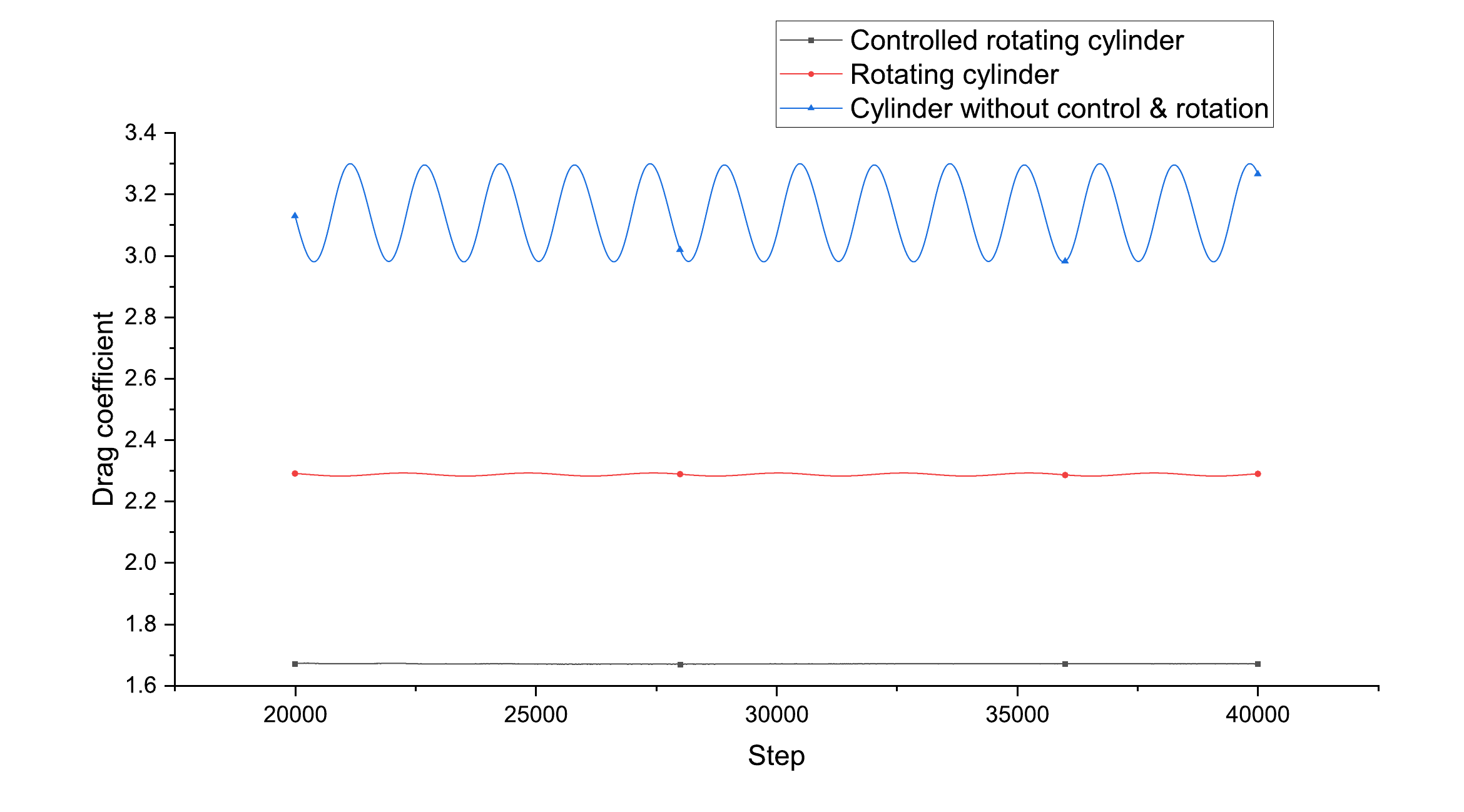}
	\caption{Drag coefficient of Re = 200 in three different configurations after the flow stabilization. \label{250}}
\end{figure}

Figure \ref{250} shows the drag coefficient of three different cases for $Re = 200$ after the flow stabilization, the first being the cylinder in the flow field without rotation or active flow control, the second graph is the case of the cylinder with 0.2 rotation and no active flow control, and lastly, the third graph shows the case with cylinder rotation of 0.2 and active flow control performed by the DRL agent. In all three cases, the best-performing parameters were used (number of actuation = 1000 and maximum allowed flow rate = 0.2).

\begin{figure}
	\centering
	\includegraphics[width=1\textwidth]{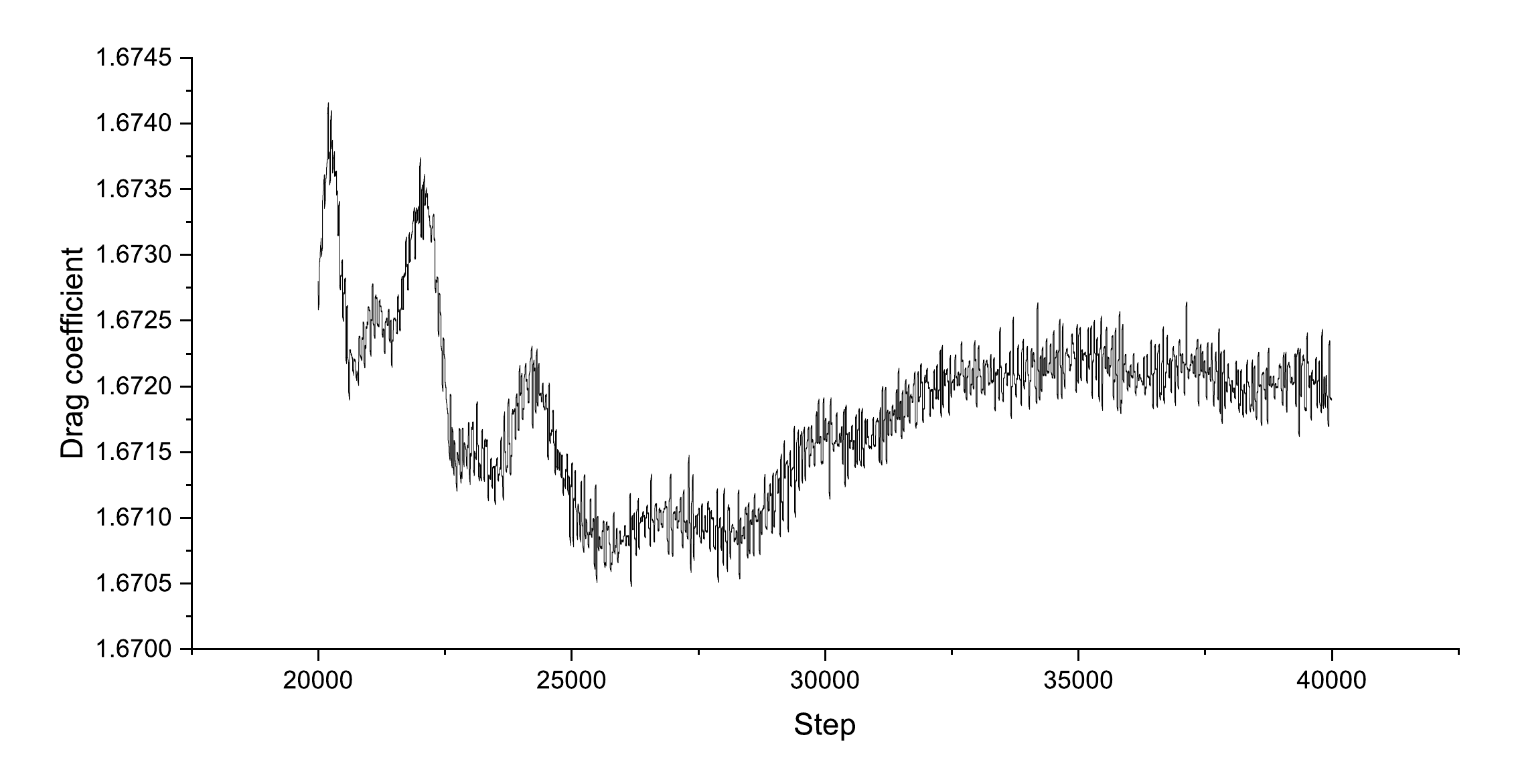}
	\caption{Drag coefficient of Re = 200 after the flow stabilization. \label{270}}
\end{figure}

\begin{figure}
	\centering
	\includegraphics[width=1\textwidth]{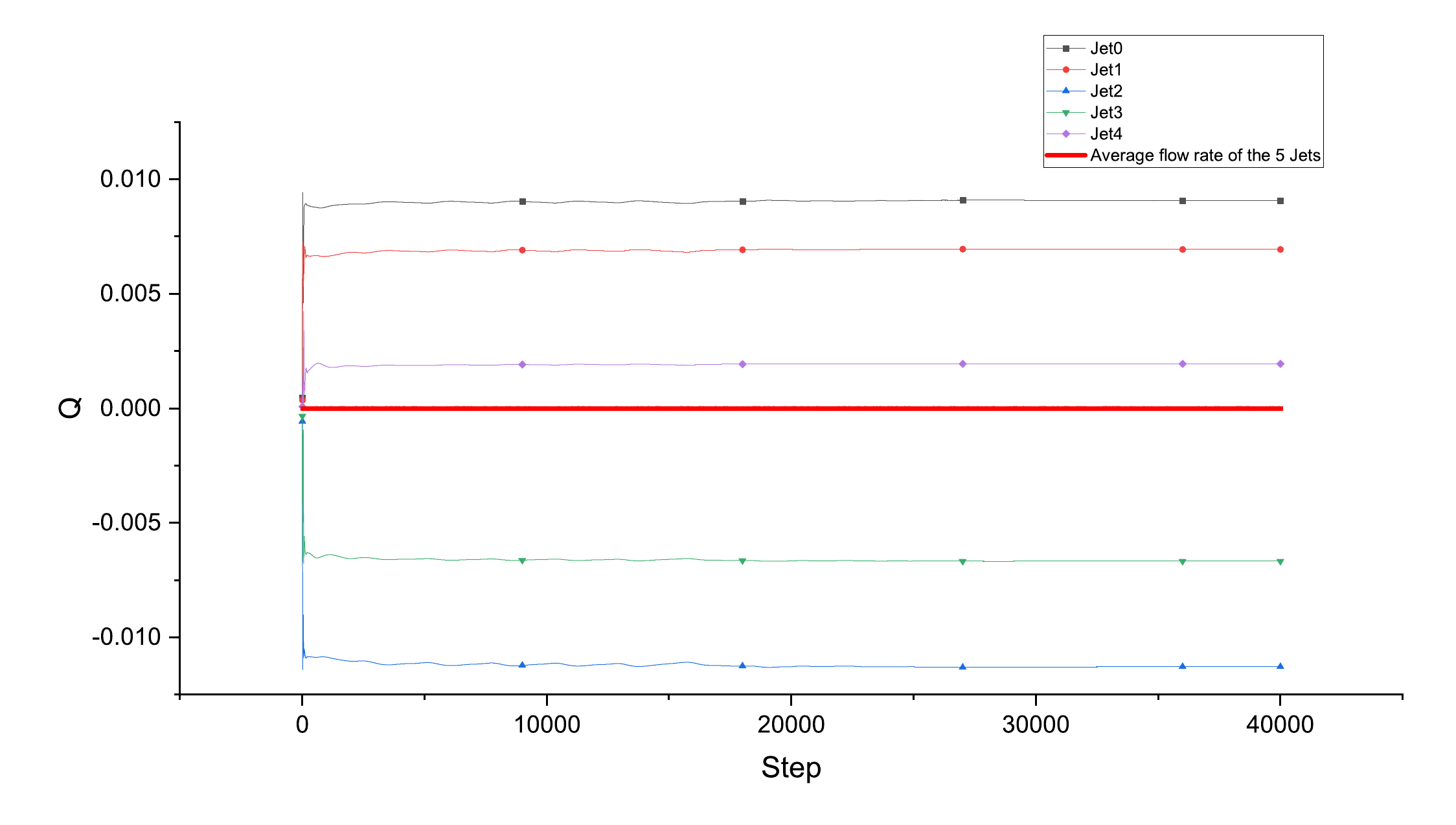}
	\caption{Jet flow rate during control. \label{290}}
\end{figure}

Figure \ref{270} shows the drag coefficient of the case with AFC and rotation, the variance of the drag coefficient in the first case (without AFC and rotation) was $0.0127$, for the second case (rotating cylinder without AFC) was $1.2075\times10^{-5}$, and finally the agent was able to further reduce the variance to only $2.56\times10^{-8}$. Also, each jet's flow rate can be seen in figure \ref{290}. 

Unfortunately, due to the hardware limitations and the exponential calculation time required for higher Reynolds numbers, this number was the limit for us. 

\section{Conclusions}
In this work, the most efficient number and position of jets, the optimum sensor number, and locations were computed. Also, rotation was added to the cylinder alongside the DRL-controlled jets, and the behavior of the agent with different controlling parameters and accesses was observed. In the end, we found the most appropriate control parameters for the rotating cylinder and the fixed case. First, we found that having more sensors at diverse locations is not always an effective choice and the sensor number and locations should be determined based on the need of the user and configuration. Also, we showed that in order to have robust learning, there is a need for at least 2100 epochs.

Secondly, the most efficient configuration of jets, based on the cases tested here and with the discussed parameters occurs when we have one jet on the opposite side of the stagnation point and four others at $45^{\circ}$, $90^{\circ}$, $270^{\circ}$, $315^{\circ}$. In this case, it was found that allowing agent to have access to higher flow rates is not appropriate as it cannot stabilize the flow. The best-performing parameters were the case with each actuation maximum allowed flow rate of one and 80 actuation per episode, which lowered the drag from a $C_{Dmax}$ of 3.2415 to 2.1050. This is equal to a considerable drop of $35.06\%$.

Thirdly, adding rotation alongside the DRL-controlled jets can lower the drag coefficient from 3.2416 to 1.6288 in the best case, which is almost a $50\%$ reduction. In this case, the vortex shedding is almost suppressed. Contrary to the previous case, giving access to higher flow rates for the agent is usually beneficial, since in this case, the constraint is partly jet outputs. This was true for configuration III with 0.2 rotation speed as in this case, configuration II could not control the flow and reduce the drag coefficient. But, this configuration could decrease the drag force at smaller rotation speeds of 0.1 and 0.08, although its performance was worse than that of configuration III. Again we believe it's partly because of the lower maximum output of 2 jets compared to the 5 and partly because of the jet at $0^{\circ}$. 

Lastly, we introduced a possible design for rotating cylinder, reduced the sensor numbers and location to a minimum, and doubled the Reynolds number in order to showcase the possibility and the performance of our best configuration and parameters for real-world and challenging applications. It should be noted that across all cases, the agent effectively lowered the lift coefficient to zero or maintained it at a lower level compared to the initial state.

\section{Declaration of Interests}
The authors report no conflict of interest.

\bibliographystyle{unsrturl}
\bibliography{references}  






\end{document}